\documentstyle[12pt,aaspp4]{article}
\def\ioz{\int_0^{z} \xi(y)}

\def\fr{\frac}
\def\rd{\right.}
\def\ld{\left.}

\def\d{\frac{d}{d \zwo}}
\def\dw{\frac{d^2}{d \zwo^2}}
\def\dt{\frac{d^3}{d \zwo^3}}
\def\df{\frac{d^4}{d \zwo^4}}

\def\pl{\partial}
\def\zo{z_{1}}
\def\zw{z_{2}}
\def\zwo{z_{21}}
\def\z{{\it{\bf  z}}}

\def\v{{\it{\bf  v}}}
\def\x{{\it{\bf  x}}}
\def\zov{{\it{\bf  z}}_{1}}
\def\zwv{{\it{\bf  z}}_{2}}
\def\zwov{{\it{\bf  z}}_{21}}

\def\dz{{\bf \nabla_z}}
\def\nx{\hat{\it{\bf  x}}}
\def\nz{\hat{\it{\bf  z}}}
\def\nn{\hat{\it{\bf  n}}}
\def\nzo{\hat{\it{\bf  z_1}}}
\def\nzw{\hat{\it{\bf  z_2}}}
\def\vp{v_P}
\def\xit{\tilde{\xi}^{R}}
\def\xito{\tilde{\xi}^{R}_1}
\def\xitw{\tilde{\xi}^{R}_2}
\def\xitt{\tilde{\xi}^{R}_3}
\def\xitf{\tilde{\xi}^{R}_4}
\def\xitz{\tilde{\xi}^{R}_0}
\def\xitn{\tilde{\xi}^{R}_n}

\def\xitzz{\tilde{\xi}^{R}_{00}}
\def\xitzw{\tilde{\xi}^{R}_{02}}
\def\xio{\bar{\xi}_1}
\def\xiw{\bar{\xi}_2}
\def\xif{\bar{\xi}_4}

\def\a{\alpha}
\def\b{\beta}

\def\lsb{\left[}
\def\rsb{\right]}
\def\lb{\left(}
\def\rb{\right)}

\def\be{\begin{equation}}
\def\ee{\end{equation}}
\def\bea{\begin{eqnarray}}
\def\eea{\end{eqnarray}}

\def\ed{\end{document}}

\begin{document}

\title{Radial Redshift Space Distortions.}
\author{Somnath Bharadwaj \footnote{Present address: Department
of Physics, IIT Kharagpur, 721 302, 
India, e-mail: somnath@phy.iitkgp.ernet.in}}
\affil{ Mehta Research Institute, Chhatnag
Road, Jhusi, Allahabad  211 019,  India\\ e-mail:
somnath@mri.ernet.in}
\begin{abstract}
The radial component of the peculiar velocities of galaxies cause
displacements in their  positions in redshift space.   We  study
the effect of the peculiar 
velocities on the linear redshift space two point correlation function.
Our analysis takes into account the radial nature of the redshift space
distortions and it highlights the limitations of the plane parallel
approximation.  We  consider the problem of determining the value of
$\b$ and the real space two point correlation function
from the linear redshift space two point correlation function. 
The inversion method proposed here takes into account the radial
nature of the redshift space distortions and can be applied to
magnitude limited redshift surveys that have only partial sky
coverage. 
\end{abstract}
\section{Introduction.}
The redshift of a galaxy has information about both its position and
the radial component of its peculiar velocity, and as a consequence
galaxy-galaxy correlations in redshift space  differ from the
correlations amongst their real positions.  Using linear theory and
the plane-parallel approximation, 
Kaiser (1987) showed that the redshift space power spectrum is the
real space power spectrum amplified by the factor $(1+ \Omega_0^{0.6}
\mu^2)^2$ where $\mu$ is the cosine of the angle between the line of
sight and the 
wave vector. He also pointed out that the anisotropy of the
redshift space power spectrum can be used  to measure the value of
$\Omega_0$.

 The plane-parallel approximation assumes that the pairs of
galaxies between which the correlation is being measured are
sufficiently far away so  that their separation subtends a very small
angle at the observer and  the displacements in redshift space
caused by their peculiar motions may be treated as parallel.  
Most of the subsequent work (Lilje \& Efstathiou 1989; McGill 1990;
Loveday et al. 1992; Hamilton 1992; Gramman, Cen \& Bahcall 1993;
Bromley 1994; Fry \& Gazta\~{n}aga 1994; Fisher et al 1994; Fisher 1995;
Cole, Fisher \& Weinberg 1994, 1995; Matsubara \& Suto 1996;) are
based on the plane-parallel approximation.

A proper analysis however requires that the radial nature of the
displacements in redshift space  is taken into account. This is
required if pairs of galaxies  
which subtends a large angle are to be also included in the analysis. 
This was first taken into account in the analysis of 
Fisher, Scharf \& Lahav (1994) who decomposed the angular behaviour of
the density field (in redshift space) into spherical harmonics and
integrated out the radial dependence after multiplying the
density field with a Gaussian radial window function. 
The power spectrum of the  coefficients of this expansion was
then used to determine the value of $\b=\Omega_0^{0.6}/b$, and they
obtained the value $\b=1.0\pm 0.3$ on applying  this technique on the
1.2-Jy {\it IRAS} redshift survey. This method was improved by  
Heavens \& Taylor (1995) who expanded the radial dependence of the
density field into spherical Bessel functions, but both these analysis
have the drawback that they require a prior knowledge of the linear
power spectrum $P(k)$. These methods were further refined by Ballinger,
Heavens \& Taylor (1995) who do not fix the shape of $P(k)$ but
allow it to vary in six bins in $k$ space. All these methods have
the limitation that they  require the galaxy survey to have full sky
coverage and they cannot be applied to two dimensional redshift surveys. 

Hamilton \& Culhane (1996), and Zaroubi \& Hoffman (1996) have
calculated the linear  two point correlation in redshift space taking
into account the radial nature of the distortions. Zaroubi \& Hoffman
(1996) have also investigated the mode-mode coupling that arises when
the analysis is done in Fourier space, but they have not addressed the
problem of determining the value of $\b$ and the real space correlation
in detail. This problem has been discussed in  detail by Hamilton
\& Culhane (1996) who propose a method for determining $\b$
in a manner which does not require any prior assumption about the real
space correlation function.  They have
studied the eigenfunctions of the 'spherical distortion operator'
which relates the real space correlation to its redshift space
counterpart. The observed redshift space correlations is
expanded in terms of these eigenfunctions and the ratio of the values
of these coefficients can be used to obtain $\b$. The problem of
dealing with the effect of the radial nature of the redshift space
distortions  on the two point correlation function has also been
considered in a recent paper by Szalay, Matsubara \&  Landy (1998). 

In this paper we have investigated the effect  of the radial nature of
the redshift space distortions on the linear two point correlation
function .  The redshift space two
point correlation is a function of the triangle formed by the
observer and the pair of galaxies for which the correlation is being
measured. In order to get a better understanding of the effects of the
redshift space distortions  we study in some detail how the redshift
space correlation function changes with the shape of the triangle, the
value of $\b$ and the slope of the real space correlation. We also
address the question as to when  the radial 
nature of the distortions is important and when they may be ignored
and the plane parallel  approximation be used instead. In section
3. we address the problem of 
determining the value of $\b$ and the real space correlation function
from the observed redshift space correlation function taking into
account the radial nature of the distortions. 
The analysis of  Hamilton \& Culhane (1996) is based on the assumption
that the selection function has a power law spatial dependence. We
have investigated whether this assumption is justified for a magnitude
limited sample. In this paper we discuss the inverse problem for two
different situations (1.) assuming that the selection function is a
power law (2.) for a more realistic form of the selection function
which can be  used in a magnitude limited survey.  Both the inversion
methods proposed here can be applied to redshift surveys that have
partial sky coverage. 

\section{The linear redshift space correlation.}
A large amount of our information about the spatial distribution of
galaxies is inferred from redshift surveys which provide
angular positions and  redshifts of  a large number of galaxies.
The distance to galaxies is very hard to measure and the
analysis of redshift surveys has to rely on the redshift as an
indicator of the distance to the galaxies. 
This has a drawback because in addition to the 
Hubble expansion the redshift  has contributions from the
radial component of the peculiar   velocity  of our  Galaxy and the 
galaxy being observed.   
The peculiar  velocity of our  Galaxy has been determined from the
dipole anisotropy observed in the CMBR  (Kogut et al. 1993)  and this
can be corrected for in all observations made from our Galaxy but  the
contribution from the radial component of the peculiar velocity of the
galaxy being observed remains in its redshift.

Using  the vector $\z$ to denote the position of a galaxy in the 
3-dimensional redshift space formed by the   angular positions and
redshifts, the relation between $\z$ and the actual position of the
galaxy $\x$   can be written as   
\begin{equation}
\z=\x + \nx (\v \cdot \nx) \,.\label{eq:a1}
\end{equation}

Here the hat  denotes a unit vector ($\nx=\nx /x$) and  $\v \cdot
\nx$  is the radial component of the peculiar velocity of
the galaxy,      The units have been chosen so that  units where the
speed of light  $c$, and the  present  value of the scale factor $a$
and its time derivative $\dot{a}$ are  all unity, 

The problem is how to use quantities measured from the
distribution of galaxies in  redshift space to draw inferences about
the actual distribution of the galaxies. In addressing this problem
it is also necessary to take into account the fact that usually
the galaxies in a redshift survey are not selected  uniformly  from
the region of space surveyed. 
 For magnitude limited surveys the selection criteria is a
function of the actual distance from the observer and it is represented
by the selection function $\phi(x)$ which gives the fraction of
galaxies  selected  in the survey as a function of the distance from
the observer.  

Taking these effects into account, the  observed  number density of
galaxies in redshift space $n^R(z)$ can be  related to the number
density of galaxies  in real space  $n(\x)=<n>(1+\delta(\x))$ and the
peculiar velocity field $\v(\x)$ as (Kaiser 1987)

\be
n^R(\z)=\Phi(z) <n> \left\{ 1+ \delta(\z) - \lsb \frac{1}{z}
\lb 2 + \frac{d \, \log (\Phi(z))} {d \, \, \, \log (z)} \rb 
+    \pl_{\z} \rsb \nz \cdot  \v(\z) \right\} \,.
\label{eq:b2}
\ee
where  $\pl_{\z}= \nz \cdot \nabla_\z$ is used to denote the
derivative  in the radial direction.

	This equation is valid at linear order in $\v$, and it has
the assumption that $v<<z$ and $\Phi(x)$ is a slowly varying
function. It is convenient to define  a function 
\be
\alpha(z)=\lb 2 + \frac{d \, \log (\Phi(z))} {d \, \, \, \log (z)} \rb
\ee
where  $\alpha(z)=2$ for a volume limited sample for which  the
selection function is a  constant.

In the linear regime, in the presence of only the  growing mode of
perturbations (Peebles 1980), it is possible express the perturbation
and the peculiar velocity in terms of a potential as
\be
\delta(\x)= \nabla^2 \psi(\x)\,\hspace{1.5cm} {\rm and} \hspace{1.5cm}
\v(\x)=-\beta  {\bf \nabla} \psi(\x) \,.  
\ee
Here $\beta=\Omega_0^{0.6}/b$ where  $\Omega_0$ is the density parameter
and $b$ is the bias  parameter which takes into account  the fact that
the galaxies may be a biased tracer of the underlying matter density
which determines the peculiar velocities. 

We use these relations in equation (\ref{eq:b2}) to write the number
density of galaxies at the point $\z$ in redshift space as 

\be
n^R(\z)=\Phi(z) <n> \left\{ 1 + \lsb  \nabla^2_{\z} + \beta \lb
\frac{\alpha(z)}{z} \pl_{\z} + \pl^2_{\z} \rb   \rsb
\psi(\z)  \right\}  \,. \label{eq:b3}
\ee

and we use this  to calculate the linear redshift space two point
correlation function $\xi^R(\zov,\zwv)$ defined as 
\be
\xi^R(\zov,\zwv)=\frac{<n^R(\zov) n^R(\zwv)>-<n^R(\zov)>< n^R(\zwv)>}
{<n^R(\zov)><n^R(\zwv)>} \,. \label{eq:a9}
\ee
where the angular brackets $<...>$  denote ensemble average. In
evaluating this we  encounter the ensemble average $<\psi(\zo)
\psi(\zw)>$. As the 
universe is statistically homogeneous and isotropic
we can define  $\phi(\zwo )  =<\psi(\zov) \psi(\zwv)>$ which  is a
function of only the magnitude of the vector  $\zwov=\zwv-\zov$.
Using this we obtain  
\be
\xi^R(\zov,\zwv)=  \lsb \nabla^2_{\zov} +\beta \lb
\frac{\alpha(\zo)}{\zo} \pl_{\zov} + \pl^2_{\zov} \rb   \rsb 
\lsb \nabla^2_{\zwv} +\beta \lb
\frac{\alpha(\zw)}{\zw} \pl_{\zwv} + \pl^2_{\zwv}
\rb    \rsb \phi(z_{21}) \,. \label{eq:b4}
\ee 
for the linear redshift space two point correlation function. 
  
This should be compared with the  real space two point correlation
function 
\be
\xi(\zov,\zwv)=<\delta(\zov)\delta(\zwv)>=\nabla^4 \phi(z_{21})  \,.
\label{eq:b5}
\ee
which depends only on $\zwo$.

Equations  (\ref{eq:b4}) and  (\ref{eq:b5}) together are equivalent to
the expression  for the linear redshift space correlation
function derived  by Hamilton and Culhane (1996). 
Equation (\ref{eq:b5}) can be inverted to relate various derivatives
of the potential $\phi$ which appear in  equation (\ref{eq:b4}) to
integrals  of $\xi(x)$ and this is described  in Appendix A.
 
The expression for the redshift space two point correlation function
presented here  is valid in the regime where $\xi(\zov,\zwv)\ll 1$. In
addition there are the restrictions that  the redshifts $\zo$ and
$\zw$ are  in a  range where they are much larger  than $\sqrt{<v^2>}$
(the $r.m.s.$ peculiar velocity) 
and where the selection function does not vary too rapidly.

Unlike the real space two point correlation $\xi (\mid \zov-\zwv\mid)$
which depends on just on the distance between  the points $\zov$ and
$\zwv$, the  redshift space counterpart
depends on the triangle formed by the observer O and the points
$\zov$ and $\zwv$, and we next investigate this behaviour in some
detail. 

 The behaviour of $\xi^R(\zov,\zwv)$ is  relatively simple in the
situation where the 
two edges of the triangle $\zov$ and $\zwv$ are made very large
keeping $\zwov$ fixed . In this limit $\nzo$ and $\nzw$ are nearly
parallel i.e. $\lim_{\zo \rightarrow \infty} \nzo=\nzw=\nn$ and the peculiar
velocities of the galaxies at $\zov$ and $\zwv$ can be treated as
being parallel. In addition, if  the selection 
function is such that $\lim_{z \rightarrow \infty} (\a(z)/z) = 0$,
then the terms involving $\a$(z) can be dropped and equation 
(\ref{eq:b4}) becomes  
\be
\xi^R(\zov,\zwv)= \lsb \nabla^2_{\zwov} + \b (\nn \cdot \nabla_{\zwov})^2
\rsb^2 \phi(z_{21})
\ee
and we have  the linear redshift space two point correlation in the
plane-parallel  approximation (PPA).  In this limit the
redshift space two point correlation depends on the length of just one
edge of the  triangle  $\zwov=\zwv-\zov$, and it depends on on the  angle
between $\zwov$ and the line of sight $\nn$. This angular dependence
introduces anisotropy in the redshift space correlation and this is
well understood in PPA (Hamilton 1992).
Here we investigate the behaviour of the redshift
space correlation  function in a more general situation where the
plane-parallel approximation cannot be applied and the  radial nature
of the distortions has to be taken into account. 
In the rest of the discussion in this section we use the value
$\alpha=2$ which corresponds to a volume limited sample where 
the selection function is a constant.

In order to separately study the dependence of $\xi^R(\zov,\zwv)$ on
the shape and the size of the triangle formed by  the observer O and
the  points  $\zov$, $\zwv$ we consider a situation where the real
space  correlation function has a power law behaviour $\xi(x) \propto
x^{-\gamma}$. In this case the  effect of changing the size of the
triangle  is very simple  $\xi^R(y \zov,y \zwv)= y^{-\gamma} \xi^R(
\zov, \zwv)$,  and the ratio
\be
w(\zov,\zwv)=\xi^R(\zov,\zwv)/\xi(z_{21}) \label{eq:w1}
\ee
depends only on the shape of the triangle. We have used the function 
$w(\zov,\zwv)$ to study how $\xi^R(\zov,\zwv)$ varies with the shape of
the triangle. 

\begin{figure}
\figurenum{1}
\plotfiddle{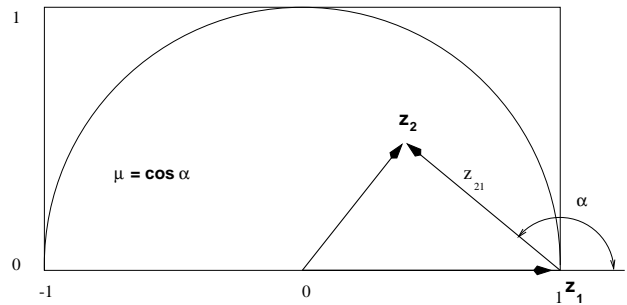}{1in}{-90}{45}{45}{-150}{120}
\caption{This figure shows the parameter space corresponding to
triangles of all possible shapes. The  observer O  at the origin 
and the point $\zov$ at (1,0) form a fixed edge of the
triangles. The third point $\zwv$ can lie anywhere in the semi-circle,
and different values of $\zwv$ corresponds to triangles of different
shapes. One possible triangle is shown in the figure.}
\label{fig:sh1}
\end{figure}

We parameterize triangles of all possible shapes by first carrying out
the following operations  which leave $w(\zov,\zwv)$ unchanged: 
(1) Label the larger of the 
two sides that originate from O as $\zov$ (2) Rotate the triangles
around O so that they all lie in the x-y plane with $\zov$ along the x
axis (3) Scale the triangles so that $\zo=1$.  At the end of these
operations, for all the  triangles $\zov$ corresponds to a
unit vector in the x direction while  $\zwv$ lies in the xy plane and
it is restricted to be  inside a  circle of unit radius  centered
around O. Triangles  which lie in the lower half  plane can be related
to triangles with the same shape in the upper half plane by reflecting
$\zwv$ on the x axis. It is thus possible to  parameterize triangles of
all   possible shapes by using the vector $\zwv$ which is restricted  
to lie inside the upper half of a circle of unit radius centered
around O. Figure \ref{fig:sh1}. shows the observer O at the point
(0,0), the point 
$\zov$ at (1,0) and the semi-circle  shows the region inside which the
point $\zwv$ must lie. Every point in the  semi-circle corresponds to a
triangle with a different shape, and one  possible triangles is
shown in  figure \ref{fig:sh1}.   We have used this parameterization to
study how  $w(\zwv)$  varies with the shape of the triangle  and the
results of this study are presented in the form of contour plots which
show contours of equal $w$ plotted at equal  intervals of $w$ for
triangles of all possible shapes. 

\begin{figure}
\figurenum{2a}
\plotfiddle{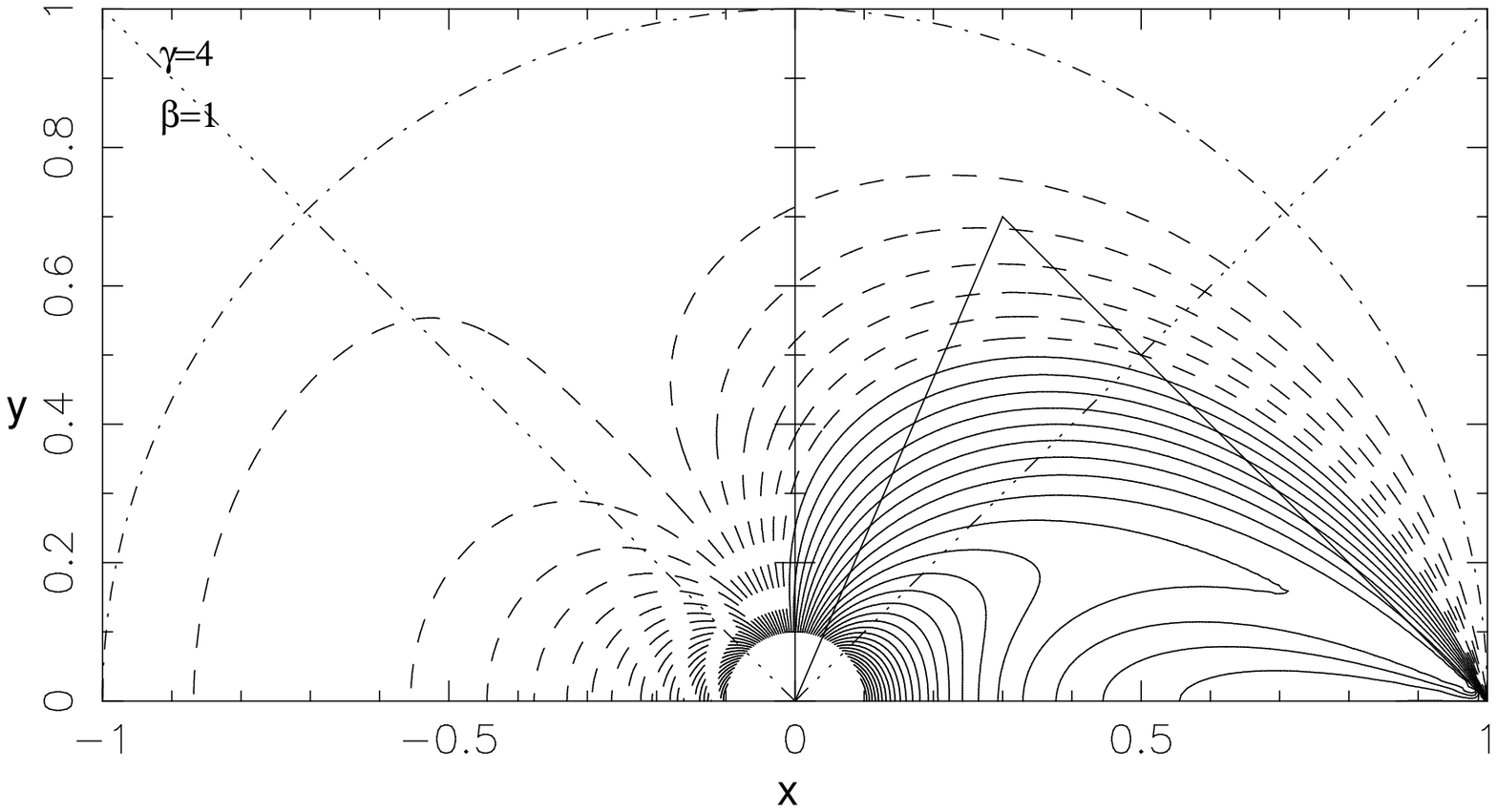}{2.5in}{0}{80}{80}{-240}{-200}
\caption{This shows contours of equal $w$ plotted in the parameter
space corresponding to triangles of all possible shapes. 
The solid lines show positive values of $w$ and the dashed lines show
negative values of $w$.} 
\label{fig:cn1}
\end{figure}

\begin{figure}
\figurenum{2b}
\plotfiddle{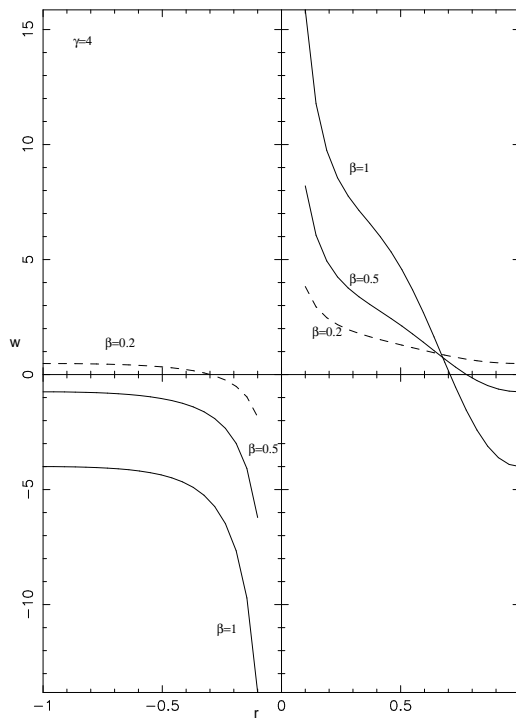}{3in}{0}{40}{40}{-80}{-30}
\caption{This shows $w$ along a cross-section at $45^{\circ}$ through figure
\ref{fig:cn1}  plotted as a function of $r$ -the distance of the 
point $\zwv$ from the observer. Positive values of $r$ correspond to
the situation where $\zwv$ and $\zov$ are in the same direction, and
$r$ is negative when they lie in opposite directions.} 
\label{fig:ct1}
\end{figure}

We have studied the behaviour of the function $w(\zwv)$ for  three
different cases $\gamma=4, 3.5$ and $2.5$ with   $\beta=1$ and the
corresponding contour plots are shown in figures \ref{fig:cn1},
\ref{fig:cn2}  and \ref{fig:cn3}  respectively.  The value of
$w(\zwv)$ along the $45^{\circ}$ cuts shown in  figures
\ref{fig:cn1}, \ref{fig:cn2}  and \ref{fig:cn3} are plotted in
figures \ref{fig:ct1}, \ref{fig:ct2}  and \ref{fig:ct3}, and
the purpose of these graphs is to show the value of $w$
corresponding to the different contours. 
Figures \ref{fig:ct1}, \ref{fig:ct2}  and \ref{fig:ct3}
also show $w(\zwv)$ for other values of $\b$ for which we have
not shown contour plots.    

Going back to the contour plots, we expect PPA to be valid in the
region near  the lower right hand 
corner of the figures where $\zwv$ is nearly equal to  $\zov$ and
$z_{21} \ll \zo$. For
comparison we have calculated $w(\zwv)$ using PPA for the case with
$\gamma=4$ and $\b=1$  for triangles of all shapes
(i.e. also beyond the region  where we would expect PPA to be valid)
and this is 
shown in figure 2.c.  A comparison of figures 2.a and 2.c shows that
as expected the figures match in the region around $\zov$, but we start
seeing considerable differences from the predictions of PPA as either
(1). the angle between $\zov$ and $\zwv$ is increased or (2). the
sizes of $\zov$ and $\zwv$ start to differ significantly. We find
that PPA  correctly describes the effect of the redshift space distortions
on the linear two point correlation function  provided
the two vectors $\zov$ and $\zwv$ do not differ  by more than
30 \% i.e. $z_{21} \le .3\, \zo$. The function $w(\zwv)$ behaves
quite differently from the predictions of PPA once the two redshift
space vectors $\zov$ and $\zwv$ differ by more than 30 \% and the
radial nature of the distortions become important in this situation. 

A comparison of figures 2.a, 3.a. and 4.a shows how the effect of the
redshift space distortion changes with the slope of the real space
correlation $\gamma$. For $\gamma=4$ and $3.5$ the redshift space
correlation is of the same sign as the real space correlation in the
forward direction (when $\zov$ and $\zwv$ are in the same direction)
and the redshift space correlation function changes sign as $\zwv$ is
turned away from $\zov$, whereas  for $\gamma=2.5$ the behaviour is
just the opposite. We also find that the shapes of the contour lines
changes significantly as the value of $\gamma$ is changed. 
For all the cases the effect of redshift space
distortions  becomes stronger as $\zwv$ approaches the observer and
$w(\zwv)$ diverges in the limit $\zw \rightarrow 0$. This behaviour is
due to the factor $\alpha(\zw)/\zw$ which appears in equation
(\ref{eq:b4}), and  the cause of the divergence can  be related to the
fact that  under the map from real space to redshift space a
non-zero volume element in real space  collapsed to a point at $z=0$. 

In figure 2.d we have shown $w(\zwv)$ for the case where $\gamma= 4$
and $\beta=0.2$. Comparing this with figure 2.a we see that a change
in the value of $\beta$ can make a qualitative difference in the
behaviour of the redshift space two point correlation
function. Although the shape of the contour lines is similar in both
the figures, the region where $w(\zwv)$ is negative differs
significantly in the two plots.

\section{Inverse Problem.}
In this section we address the problem of using the linear redshift space
correlation function measured  from redshift surveys to determine 
the value of $\b$ and the real space correlation function.

The inverse problem is very easily solved if $\xi^R(\zov,\zwv)$ is
restricted to a region where PPA is valid. The  function
$\xi^R(\zov,\zwv)$ can then be written as a function of 
$\zwo=\mid \zwv-\zov \mid$ and $\mu$  which is the cosine of the
angle (shown in figure \ref{fig:sh1} ) between $\zwov$ and $\nn$ - the
line of sight to the pair of 
galaxies.  In PPA the angular dependence of $\xi^R(z_{21},\mu)$ is
very 
simple and it can be expressed in terms of  Legendre polynomials 
$P_l(\mu)$ as
\be
\xi^R(\zwo,\mu)=\sum_{l=0}^{\infty} P_l(\mu) \xi^R_l(\zwo)
\label{eq:c1} \,
\ee  
where the  $\xi^R_l(\zwo)$s are different angular moments of
$\xi^R(\zwo,\mu)$. As shown by Hamilton 
(1992), only the first three even moments have nonzero  values and
the value of $\b$ can be obtained from the ratios of these angular
moments   determined from redshift surveys.

The situation is changed if the radial nature of the redshift space
distortion is taken into account as the behaviour of
$\xi^R(\zov,\zwv)$ now  depends on the shape of the triangle  formed by 
$\zov, \zwv$ and O.  This  has been studied by Hamilton \& Culhane
(1996) who have expressed the shape dependence of $\xi^R(\zov,\zwv)$
in terms of five shape functions and they have proposed a method for
measuring the value of $\b$ based on  this.
In their work Hamilton and Culhane (1996) have made the simplifying
assumption that 
the selection function can be described by a power law $\Phi(z)
\propto z^{\alpha -2}$ which implies that $\alpha(z)=\alpha$ is a
constant.

Here we propose a different method for analyzing the linear redshift
space two point correlation function  and determining the value of
$\beta$. In the first part of the analysis presented below we assume
that $\alpha(z)=\alpha$ is a constant and later on in a separate
subsection we treat  the inverse problem  for a more
realistic selection function. 

  In this analysis  $\zov$ and $\zwv$, the two
sides of the triangle which originate from O,  are not treated on a equal
footing. Writing one of the sides (say $\zov$) as   $\zov= \zo \nn$,
all possible triangles can be parameterized using the lengths of two of
the sides -  $\zo$ and $\zwo$,  and  $\mu$ the cosine of the angle
(shown in figure \ref{fig:sh1}) between $\zwov$ and  $\nn$.  This way
of parameterizing is very  similar to that used in 
PPA,  except that in PPA $\nn$ is the common direction along which
both  $\zov$ and $\zwv$ lie whereas now $\nn$ refers to the direction
of one of the sides $\zov$. Also, in PPA two parameters $\zwo$ and
$\mu$ suffice to describe the behaviour of $\xi^R$, whereas we require
an additional  third parameter  $\zo$ if the radial nature of the
distortions are taken into account. For a fixed value of $\zwo$,  PPA
corresponds to the limit $\zo \rightarrow \infty$, and hence the value
of $\zo$ does not appear in $\xi^R$ in PPA.  

It is convenient to express $\zo$ 
in terms of the dimensionless ratio 
\be
s=\frac{\zwo}{\zo}
\ee
and we use $(s, \zwo,\mu)$  to parameterize $\xi^R$. For a fixed value
of $\zwo$,  the limit $s \rightarrow 0$ corresponds
to PPA and  the effects of the radial nature of the redshift space
distortions become more important as  $s$ increases.  
 
The  set of parameters $(s,\zwo,\mu)$ has the feature that
any triangle can be described by  two sets of values of 
$s$ and $\mu$. For example for a triangle formed by a pair of galaxies
(which we call A and B) and O, we get different values of $s$ and
$\mu$ depending on whether we label A as  $\zov$ or we label B as
$\zov$. Here we take the point of view that both the labellings should be
used and as a consequence each pair of galaxies contributes to
$\xi^(s,\zwo,\mu)$ for two different values of $s$ and $\mu$.

We first consider the angular dependence of $\xi^R(s,\zwo,\mu)$, and
following the analysis used in PPA (equation \ref{eq:c1}) we decompose
$\xi^R$ in terms of  Legendre polynomials. We find that unlike in PPA,
now all the moments have non-zero values  and this does not provide a
convenient way of analyzing $\xi^R$. The analysis becomes
considerably  simpler if we use the redshift weighted 
correlation function $\xit$ defined as
\be
\xit(s,\zwo,\mu)= \frac{\zw^2}{\zo^2} \,  \xi^R(s,\zwo,\mu)\,.
\ee
where $(\zw/\zo)^2$ can be written as $(1 + s^2 + 2 s \mu)$. 
The angular dependence of $\xit(s,\zwo,\mu)$ is much
simpler, and expanding it in terms of Legendre polynomials  
\be
\xit(s,\zwo,\mu)=\sum_{n=0}^{\infty} \xit_n (s,\zwo) P_n(\mu)
\label{eq:cc1} 
\ee
we find that only the first five moments have non-zero values  which
are shown  below 
\bea
\xitz(s,\zwo)&=& \lb 1 + \frac{2}{3} \b + \frac{1}{5} \b^2 \rb \xi(\zwo) 
+ s^2 \lsb (1 + \frac{4}{3} \b + \frac{1}{3} \b^2 ) \xi(\zwo) \right.
\nonumber \\ &-& \left.  \frac{1}{6} \a^2 \b^2 \xio(\zwo) + \frac{1}{9} \b
(-6+\a -2 \b - \a \b + \a^2 \b) \xiw(\zwo) \rsb \label{eq:c2} \\ \nonumber \\ 
\xito(s,\zwo)&=& s \lsb \lb 2 + \frac{16}{5} \b + \frac{6}{5} \b^2 \rb
\xi(\zwo)  
-  \lb \frac{28}{15} \b  + \frac{4}{5} \b^2  \rb \xiw(\zwo) \rsb
\nonumber \\
&+& s^3 \a \b  \lsb - \frac{1}{6} \a \b \xio(\zwo) - \frac{1}{3} \xiw(\zwo)
- \frac{1}{15}\b(3-\a) \xif(\zwo) \rsb  \label{eq:c3}
\\ \nonumber \\
\xitw(s,\zwo)&=&  \lb \frac{4}{3} \b  + \frac{4}{7} \b^2 \rb  \lb \xi(\zwo)
- \xiw(\zwo) \rb + s^2 \lsb \frac{2}{3} \b ( 1+ \b) \xi(\zwo) \right.
\label{eq:c4} \\ &+& \left.
\frac{1}{9} \b \lb -6 -4 \a +2 \b + \a \b - \a^2 \b \rb \xiw(\zwo) -
\frac{1}{5} \b^2 \lb 4 + \a - \frac{1}{3} \a^2 \rb \xif(\zwo) \rsb 
\nonumber  \\ \nonumber \\
\xitt(s,\zwo)&=&  s \frac{4}{5} \b \lsb \lb 1 + \b \rb \xi(\zwo) -
(1-\b) \xiw(\zwo) - 2 \b \xif(\zwo) \rsb \label{eq:c5}
\\ \nonumber \\
\xitf(s,\zwo)=&=& \frac{4}{35} \b^2 \lb 2 \xi(\zwo) + 5 \xiw(\zwo) - 7
\xif(\zwo) \rb \,. \label{eq:c6}
\eea

We have used the relations given in the Appendix A to express the
derivatives of the potential $\phi(\zwo)$ in terms of volume
integrals of the real space two point  correlation function  
${\bar{\xi}_n}$  which are defined as 
\be
\xio(\zwo)=\frac{2}{\zwo^2} \int_{\infty}^{\zwo} \xi(r) r dr 
\label{eq:cc2}
\ee
and for $n >1$
\be
\bar{\xi}_n(\zwo)=\frac{n+1}{\zwo^{n+1}} \int_0^{\zwo} \xi(r) r^n dr \,,
\label{eq:cc3}
\ee  

The first point to note is that for a fixed value of $\zwo$,  
in the limit $s \rightarrow 0$ we recover  the results calculated by
Hamilton (1992) in the plane parallel approximation. As expected,  the
odd moments all vanish in this limit. 

The effect of including the radial nature of the distortions manifests
itself through the $s$ dependent terms.  The different
angular moments all have a very simple dependence on $s$ and this
involves  at most a cubic polynomial in $s$.  The
angular 
moment $\xitf$ has no $s$ dependence, and the expression  calculated
for this moment using PPA remains unchanged if the radial nature of
the distortions is taken into account. 

The procedure for determining  the value of $\b$ from redshift surveys
is now quite straightforward in principle. 
The first step is to estimate  the redshift  weighted correlation
function  $\xit(s,\zwo,\mu)$   using all the  pairs of  galaxies  in
the survey. For a fixed value of $\zwo$, for a survey  which  extends from
a  redshift $z_a$ to a redshift  $z_b$, the variable $s$ will lie
in the range  $\zwo/z_a \ge  s \ge  \zwo/z_b$.

The second step is to  decompose the angular dependence of the
observationally determined $\xit(s,\zwo,\mu)$ in terms of Legendre 
polynomial. This is possible only for  those values of $(s,\zwo)$
where  there are observations of $\xit(s,\zwo,\mu)$ for both positive
and negative $\mu$, and in general the  range of $(s,\zwo)$ where this
is possible will depend on the geometry of the redshift survey. 
The angular decomposition can be   used to obtain  estimates for the
first five  angular  moments $\xitz(s,\zwo)$ to $\xitf(s,\zwo)$.  We
expect all the higher moments  to be zero in the linear regime.

For a fixed value of $\zwo$, the $s$  dependence of the angular moments
$\xit_n(s,\zwo)$  is very simple in the linear regime.
The third step is to do a least squares fit for the $s$ dependence of
the $\xit_n(s,\zwo)$s using  polynomials in $s$ of the form  predicted
by equations (\ref{eq:c2}) to  (\ref{eq:c6}). For example,  the
monopole determined from redshift surveys can be fitted  using a
quadratic function of the form $\xitz(s,\zwo)=\xitzz(\zwo)\, s^0 +
\xitzw(\zwo)\, s^2$, where  $\xitzz(\zwo)$ and $\xitzw$ are the unknown
quantities  which have to be determined from the fit.

The coefficient of the terms $s^0$ in the fits for the even moments
gives $\xitz(0,\zwo)$, $\xitw(\o,\zwo)$ and  $\xitf(s,\zwo)$ which
correspond to the values of these moments in PPA.
Effectively, this allows us to use  pairs of galaxies for which the
radial nature  of the redshift space distortion is important (large
values of $s$) to predict the behaviour of the angular moments at
$s=0$ which corresponds to PPA. Once these are known we can use the
method  proposed by  Hamilton (1992) for determining $\b$ and
$\xi(\zwo)$  using results calculated in the  plane-parallel
approximation.   

There are various possible ways of doing this and one way
is to use the relations
\bea
(1+\frac{2}{3} \b + \frac{1}{5} \b^2) \lsb\xi(\zwo) - \xiw(\zwo)\rsb &=& 
\xitz(0,\zwo) - \frac{3}{\zwo^3} \int_0^{\zwo} \xitz(0,r) r^2 dr \label{eq:c7}
\\
(\frac{4}{3} \b + \frac{4}{7} \b^2 ) \lsb\xi(\zwo) - \xiw(\zwo)\rsb
&=& \xitw(0,\zwo)\label{eq:c8} \,.
\eea 
The right hand side of equations (\ref{eq:c7}) and  (\ref{eq:c8})
involve quantities which can be determined from the redshift surveys
by using the procedure discussed above.
The ratio of these two equations can then be used to  calculate
$\b$, which can in turn be used to determine $\xi(x)$.

\subsection{A Realistic Selection Function.} 
Until now we have discussed the inverse problem for a situation where
the selection function has a specific form $\Phi(z) \propto z^{\a-2}$
which implies that $\a(z)=\a$ is a constant. 
In this subsection we consider a realistic situation where we
have a magnitude limited survey. We first  investigate the validity of
the assumption that $\a(z)$ is a constant.  

For a magnitude limited sample with a lower apparent magnitude limit
$m_{min}$ and an upper magnitude limit $m_{max}$, the selection
function is related to $N(M)$ the differential galaxy luminosity
function  by
\be
\Phi(z)=C_1  \int^{M_{max}(z)}_{M_{min}(z)} N(M) dM \label{eq:f1}
\ee
where the limits in the integral can be related to the 
limits in the apparent magnitude using the relation between absolute
and apparent magnitudes, and  the galaxy luminosity function $N(M)$ is
usually modeled using the Schechter function (Schechter 1976). 
The normalization constant $C_1$ in equation (\ref{eq:f1}) is of no
interest in this discussion as it does not affect $\a(z)$, and
it is only the shape of the  selection function which is of interest.
 
As an example we have used the apparent magnitude limits and the
luminosity function given for the N112 subsample of the Las Campanas
Redshift Survey - LCRS (Lin et. al. 1996) to calculate  $\Phi(z)$ and
$\a(z)$ (shown in figures \ref{fig:f1} and \ref{fig:f2}) for $q_0=.5$
and $H_0=100\, {\rm km/s/Mpc}$ . We have set the normalization $C_1$ to
an arbitrary number chosen  for the  convenience of  plotting $\Phi(z)$. 

We find that the assumption that $\a(z)$ is a constant does not
correctly describe the behaviour of the calculated values of $\a(z)$ 
shown in figure (\ref{fig:f2}). The behaviour of $\a(z)$ in the
redshift range  $0.05 \le z \le 0.19$ can be fitted  using  a function
of the form  $\a(z)=\a - z^2/q^2$ with $\a=1.7$ and
$q=0.072$. This corresponds to a selection  function of the 
form $\Phi(z)=C z ^{\a-2} e^{-z^2/2 q^2}$ (shown in s figure
\ref{fig:f2})  and we find that this fits the calculated selection
function to better than $2 \%$ in the range  $0.05 \le z \le 0.19$.
Figure (\ref{fig:f2}) also shows $n(<z)$ - the fraction of the
galaxies we  expect to find at  redshifts less than  $z$, and  we see
that about $90 \%$ of the  galaxies are expected to lie in the 
redshift range over which this fit  is valid.

In the rest of this paper we assume  that  the selection function
is  of the form 
\be
\Phi(z)=C z^{\a-2} e^{-z^2/2 q^2} \label{eq:f2}
\ee
with
\be 
\a(z)=\a -  z^2/q^2 \label{eq:f3} \, 
\ee 
where $\a$ and $q$ are constants which have to be determined for
the particular galaxy survey being considered. We expect the  selection
function to behave as predicted by equation (\ref{eq:f2}) for in any
magnitude limited survey where the  luminosity function is of the
Schechter form which is a product of a power law and an exponential.  
It may also  be noted that   a fit of the form  $\a(z)=-z^2/q^2$ which
corresponds to a  selection function $\Phi(z)=C  e^{-z^2/2q^2}$  is
reasonably accurate $(\sim 10 \%)$ and this may also be used instead of
equation (\ref{eq:f2}).

The form of the selection function in equation (\ref{eq:f2})
introduces a new length-scale $q$  and $\xi^R(\zov,\zwv)$ now behaves
quite 
differently from the case where $\a(z)$ is a constant. The most
important  difference is that now the limit $\lim_{z \rightarrow
\infty}  \a(z)/z$ diverges and hence $\xi^R(\zov,\zwv)$ as given by
equation (\ref{eq:b4}) diverges in the plane parallel approximation
which corresponds to  the limit $\zov \rightarrow \infty$ and    $\zwv
\rightarrow \infty$ with $\zwov$ held fixed.  Before proceeding
further it is necessary to clarify the fact that this divergence in
the behaviour of  $\xi^R(\zov,\zwv)$ is not a physical effect and we
do not expect to find a divergent behaviour in the $\xi^R(\zov,\zwv)$
determined from redshift surveys  when $\zov$ or $\zwv$ are made very
large. The  divergent behaviour predicted by equation (\ref{eq:b4})
arises because the derivation of  this equation involves Taylor
expanding $\Phi(z+\vp)$ ($\vp$ is the radial component of the peculiar
velocity) in powers of $\vp$. For the selection function
given in equation (\ref{eq:f2}) this involves expanding the function
$e^{-(z+\vp)^2/2 q^2}=e^{-(z^2+2 z \vp +\vp^2)/2 q^2}$ in powers of
$\vp$. Such an expansion is valid only if $z\, \vp \ll q^2$ and $\vp
\ll q$, and the expansion is invalid if we take the limit $z
\rightarrow \infty$. As a consequence equation (\ref{eq:b4}) also
becomes invalid for very large values of $\zov$ or $\zwv$, and  
this gives rise to a divergent behaviour in $\xi^R(\zov,\zwv)$. 
Equation  (\ref{eq:b4}) is valid provided $\sqrt{<v^2>} \zo/q^2 \ll 1$,
$\sqrt{<v^2>} \zw/q^2 \ll 1$ and $\xi(\zwo)$ is in the linear
regime, and it does not correctly describe the behaviour of 
$\xi^R$ if any of these conditions are not satisfied.  

We next check over what redshift range  equation (\ref{eq:b4}) is
valid for LCRS.  Using the values $\sqrt{<v^2>} \sim 0.004$ and $q =
0.07$,   we find that  equation (\ref{eq:b4}) can be used  at
linear  scales  over the entire redshift range for which the fit given
by equation (\ref{eq:f2}) is valid. Having ascertained the fact that
equations (\ref{eq:b4}) and (\ref{eq:f2}) are both valid over the
redshift range in which most of the galaxies lie, we proceed to
investigate the inverse problem for a situation where the linear
$\xi^R$ is described by these equations.

As a consequence  of the divergent behaviour discussed above, it  is
not possible to extrapolate the $\xi^R$ determined from pairs of
galaxies for which equations (\ref{eq:b4}) and (\ref{eq:f2}) are
valid  to  obtain $\xi^R$ 
in the plain parallel approximation. Hence, it is not possible
to obtain the value of $\b$ using the  method  discussed earlier for
the case where $\a(z)$  is a constant,  and a different inversion
method has to  be used for the selection function given by  equation
(\ref{eq:f2}). 

	We find that in this case it is convenient to use
$(\zo,\zwo,\mu)$ to  parameterize the redshift weighted two point
correlation function 
\be
\xit(\zo,\zwo,\mu)=\frac{\zw^2}{\zo^2} \xi^R(\zo,\zwo,\mu)
\ee
where $\zw$ can be written as $\zw^2 =\zo^2+\zwo^2+2 \zo \zwo \mu$. 

 Decomposing  $\xit(\zo,\zwo,\mu)$ in terms of Legendre  Polynomials
(equation \ref{eq:cc1})  we again find that only the first five
moments are  non-zero. In 
this case it is more convenient to work  directly in terms of the
potential $\phi(\zwo)$ instead of expressing the derivatives of
$\phi(\zwo)$  in terms of volume averages of $\xi(\zwo)$.  The
expressions for the  angular moments $\xitt$ and $\xitf$ are given
below  
\bea
\xitt(\zo,\zwo)&=&\lsb {{4\,\b\,\zwo}\over {5\,{\it \zo}}} 
\lb 1+ \b \rb \, {\df}  +
\lb {{8\,\b}\over {5\,{\it \zo}}} - 
     {{6\,{\b^2}\,{\zwo^2}}\over {5\,{q^2}\,{\it \zo}}} \rb {\dt} 
\right.  \nonumber \\  \nonumber \\ &+& \left. 
\lb {{4\,\alpha \,{\b^2}\,\zwo}\over {5\,{q^2}\,{\it \zo}}} - 
     {{4\,{\b^2}\,\zwo\,{\it \zo}}\over {5\,{q^4}}} \rb   {\dw}\rsb
\phi(\zwo)  \label{eq:f4} \\ \nonumber \\
\xitf(\zwo)&=&{\b^2} 
{{8}\over {35}}  \lsb \,{\d}  - {{2\,\zwo}\over{{q^2}}}\rsb \,{\dt}
\phi(\zwo) \label{eq:f5}
\eea
The expressions for the other non-zero $\xitn$s are very similar to
equation (\ref{eq:f4}) and these are presented in Appendix B. 
The equation for $\xitf(\zo,\zwo)$ is much simpler than the equations
for the other angular moments and it does not involve $\zo$. This
equation can be integrated once  to obtain  $ d^3\phi(\zwo)/d \zwo^3$
in terms of the $\xitf(\zwo)$  determined  from 
the redshift survey, and this gives us the relation 
\be
\dt \phi(\zwo) = \frac{35}{\b^2 8} e^{\zwo^2/q^2} 
 \int^{\zwo}_{z_L} \, dy \, e^{-y^2/q^2} \xitf(y)   +
e^{(\zwo^2-z_L^2)/q^2} A_1  \label{eq:f6}
\ee
where we only use the values of $\xitf(\zwo)$ at separations  $\zwo
\geq z_L$  where we expect the correlations to be in the linear
regime. The constant of integration $A_1$ corresponds to 
$ d^2 \phi(\zwo)/d \zwo^2$ at the point $\zwo=z_L$. Equation
(\ref{eq:f6}) can be integrated once more to obtain 
\be
\dw \phi(\zwo) = \frac{35}{\b^2 8}  \int^{\zwo}_{z_L} dr \,  e^{r^2/q^2} 
 \int^{r}_{z_L} dy \, e^{-y^2/q^2} \xitf(y)   +  \int^{\zwo}_{z_L} dr           e^{(r^2-z_L^2)/q^2} A_1 + A_2 \label{eq:f7}
\ee
where $A_2$ is a constant of integration which corresponds to the value
of $d^2 \phi(\zwo)/d \zwo^2$ at $\zwo=z_L$.  Equation (\ref{eq:f6})
can also be used to obtain  the relation
\be
 \df \phi(\zwo)= \frac{35}{\b^2 8} \xitf(\zwo) + \frac{2 \zwo}{q^2}
\dt \phi(\zwo) \,. \label{eq:f8}
\ee

	These equations (\ref{eq:f6}), (\ref{eq:f7}) and (\ref{eq:f8})
allow us to determine the value of the derivatives of the potential
$\phi(\zwo)$ in terms of the observed $\xitf(\zwo)$ and three unknown
parameters $\b$, $A_1$ and $A_2$. Using these relations in equation 
(\ref{eq:f4}) we obtain the expression  for $\xitt(\zo,\zwo)$
presented below 
\bea
\xitt(\zo,\zwo)&=&  {{7\,\zwo}\over {2\, \zo}} \lb 1 +  {1 \over \b} \rb
 \,\lsb \xitf(\zwo) \rsb 
\nonumber \\ \nonumber \\  &+& 	
{7\over {\zo}} \lb {1 \over \b} + {{\zwo^2} \over {4\,{q^2}}} +  
{{\zwo^2}\over {\b\,q^2}}  \rb \,  \lsb e^{\zwo^2/q^2} 
 \int^{\zwo}_{z_L} \, dy \, e^{-y^2/q^2} \xitf(y) \rsb
   \nonumber \\ \nonumber \\  &+&
{{7 \, \zwo}\over {2\,q^2}} \lb {{\alpha} \over {\zo}} - 
{{\zo}\over {q^2}} \rb  \, \lsb \int^{\zwo}_{z_L} dr \,
e^{r^2/q^2}   \int^{r}_{z_L} dy \, e^{-y^2/q^2} \xitf(y) \rsb
   \nonumber \\ \nonumber \\  &+&
{{4 \, \b^2 \zwo} \over {5\, q^2}} \lb {{\alpha} \over {\zo}} -
{{\zo}\over {q^2}}  \rb \, \lb A_1 \, \int^{\zwo}_{z_L} dr
e^{(r^2-z_L^2)/q^2} + A_2 \rb
\nonumber \\ \nonumber \\  &+&
{{2 \b} \over {5 \zo}} \lb 4 +  {x^2 \over q^2} \lb 4+\b \rb \rb A_2 \,
e^{(\zwo^2-z_L^2)/q^2} \label{eq:f9}
\eea
This equation allows us to use the $\xitf(\zwo)$ determined from 
observations to predict the values of $\xitt(\zo,\zwo)$.  The
quantities which involve the observed values of $\xitf(\zo,\zwo)$ have
been enclosed in  square brackets in the above equation. The relation
between the observed $\xitf$ and the predicted values for $\xitt$
also involves three  parameters  $\b$, $A_1$  and $A_2$ which are 
the only  unknown quantities in equation (\ref{eq:f9}). 
By comparing the values  of $\xitt(\zo,\zwo)$  predicted  by equation
(\ref{eq:f9}) with the  observed values of  $\xitt(\zo,\zwo)$ it is
possible to determine the value of these parameters  $\b$, $A_1$  and
$A_2$  for which the predicted $\xitt$ best fits the observed $\xitt$.

A similar procedure can also be carried out using the angular moments
$\xitz$, $\xito$ and $\xitw$, and the expressions for these quantities
in terms of the observed $\xitf$ and the parameters $\b$, $A_1$ and
$A_2$ are presented in appendix B.  The best fitting values of the
parameters  $\b$, $A_1$  and $A_2$ determined using the  four
different angular moments should be consistent and this provides a way
of checking the validity of the method proposed here. 

One the values of the parameters  $\b$, $A_1$  and $A_2$ are known it
is quite straight forward to determine the real space correlation
$\xi(\zwo) = \nabla^4 \phi(\zwo)$. 

\section{Summary and Discussion.}
We have studied  how the peculiar velocities distort the linear two
point correlation function in redshift  space. Our analysis takes into
account the radial nature of the 
distortion. We have compared this with the linear two point
correlation calculated in the plane parallel approximation and we find
that there are significant differences in the behaviour of
$\xi^R(\zov,\zwv)$  if the two redshift space vectors  $\zov$ and
$\zwv$ differ by  more  than 30 \%. The effect of the radial nature of
the redshift space distortions become important when
either the angle between $\zov$ and $\zwv$ becomes large, or  the 
lengths of $\zo$ and $\zw$ differ significantly and the plane parallel
approximation does not correctly describe the effect of redshift space
distortions under these curcumstances.

 We have also  addressed the problem of extracting the value of
$\beta$ and the real space correlation function from the redshift
space correlation function taking into account the radial nature of
the distortions. This problem has been studied earlier by   Hamilton
\& Culhane (1996) who have assumed that $\a(z)$ is a constant. We have
tested the validity of this assumption for a magnitude limited survey and
we find that such an assumption is not justified. The inversion scheme
proposed by Hamilton \& Culhane (1996) can be applied only to  volume
limited samples where  $\a(z)=2$.

 In the first part of our analysis we have  followed Hamilton \&
Culhane (1996) in assuming that $\a(z)$ is a constant.
The inversion procedure proposed here is quite different 
from that proposed by Hamilton \& Culhane (1996). The main difference
is that we propose the use of a redshift weighted two point
correlation function $\xit$  instead of the correlation
function $\xi^R$. The  function $\xit$ has the advantage that if we
decompose its angular dependence in terms of Legendre polynomials
only the first five  angular moments are  non-zero, and all the higher
angular moments are zero.  This is not the case if we use $\xi^R$
instead.  The expressions we obtain for the  angular moments of $\xit$
are very closely related to the results one gets in the plane parallel
approximation and this makes the proposed inversion procedure very
simple. 

For a magnitude limited sample the selection function can be very well
approximated by $\Phi(z)=C z^{\a-2} e^{-z^2/2 q^2}$ and we have
analyzed the inverse problem for such a situation. We find that
in this situation also the redshift weighted correlation function has
only five non-zero angular moments, and the procedure proposed for
determining $\b$ in section 3.1 is based on this. This procedure
can be applied to magnitude limited samples.  It also has the feature
that it uses the values of the redshift space correlation only at
separations larger than some separation $z_L$ which can be chosen so
that  all scales larger than $z_L$ are in the linear regime. This
has the advantage that the inversion procedure uses only the values
of the redshift space correlation function from scales which are
definitely in the linear regime and the inversion procedure is not
affected in anyway by the non-linear scales.

Both the inversion schemes discussed in this paper can be applied to
redshift surveys with partial sky coverage - for example they can be
applied to the Las Campanas Redshift Survey where the observations are
restricted to six thin conical slices.

\appendix
\section{Appendix A.}
The equation 
\be
\xi(z)=  \nabla^4 \phi(z)
\ee
can be inverted to obtain 
\bea
\pl_i \pl_j \phi(z) = & & \frac{1}{3} \delta_{i j} (\ioz y
dy + C)- \frac{1}{2} \pl_i  \pl_j (z) \ioz y^2 dy  \nonumber \\
&-& \frac{1}{6} \pl_{i} \pl_{j}(\frac{1}{z}) \ioz y^4 dy  \,.  
\eea
where $z_i$  refers to the   components of $\z$, $\pl_i$ refers to 
the components of $\dz$ and    $C$ is a  constant whose value is fixed
by the boundary conditions. It is most natural to choose the condition 
\be
\lim_{z \rightarrow \infty} \pl_{i} \pl_{j} \phi(z) =0
\ee
which gives us 
\be 
C=- \int_0^{\infty} \xi(y) y dy \,.
\ee
We also have 
\bea
\pl_{i} \pl_{j} \pl_{k} \phi(z) = 
- \frac{1}{2} \pl_{i} \pl_{j} \pl_{k}(z) \ioz y^2 dy -  \frac{1}{6}
\pl_{i} \pl_{j} \pl_{k}(\frac{1}{z}) \ioz y^4 dy  
\eea
and 
\bea
\pl_{i} \pl_{j} \pl_{k} \pl_{l} \phi(z) &=& 
 \frac{ z_{i} z_{j} z_{k} z_{l}}{z^4} \xi(z)
\hspace{.5in}- \frac{1}{2} \pl_{i} \pl_{j}\pl_{k} \pl_{l}(z)
\ioz y^2 dy  \nonumber \\   &-& 
\frac{1}{6}  \pl_{i} \pl_{j} \pl_{k} \pl_{l}(\frac{1}{z}) \ioz
y^4 dy 
\eea
\section{Appendix B.}
Here we present the expressions for the angular moments
$\xitz(\zo,\zwo)$,  $\xito(\zo,\zwo)$ and $\xitw(\zo,\zwo)$ of the
redshift weighted linear two point correlation function
$\xit(\zo,\zwo,\mu)$ calculated for the selection function described
by equation (\ref{eq:f2}). 
\bea 
\xitz(\zo,\zwo)&=&  \lsb 1 + \fr{2 b}{3} + \fr{b^2}{5} +
\fr{\zwo^2}{\zo^2} \lb 1 + \fr{4 b}{3} + \fr{b^2}{3} \rb \rsb
\, \df \phi(\zwo) \, 
+ \, \lsb \fr{4}{3 \zwo} \lb 3 + \b \rb 
\rd \nonumber \\ \nonumber \\ &-&  \ld 
\fr{b \zwo}{15 q^2} \lb 15 + \b \rb   + \fr{\zwo}{3  \zo^2} \lb 12 +
8 \b + \a \b - \a \b^2 \rb  - \fr{\b \zwo^3}{3 q^2 \zo^2} \lb 3 + \b
\rb \rsb \dt \phi(\zwo)
\nonumber \\ \nonumber \\ &+& 
\lsb \fr{2\, \b}{3\, q^2}  \lb\a \b -3 \rb + \fr{4}{\zwo^2}  +
\fr{1}{3 \zo^2} \lb 12 + 2 \a \b - \a^2 \b^2 \rb
\rd \nonumber \\\nonumber \\ &+& \ld 
\fr{\b \zwo^2}{q^2
\zo^2} \lb \a \b -2 \rb -\fr{b^2}{3 q^2} \lb 3 \zwo^2 + \zo^2 \rb
\rsb \, \dw \phi(\zwo) \label{eq:ap1}
\eea
\bea
\xito(\zo,\zwo)&=& \lsb \fr{2\, \zwo}{5\, \zo} \lb 5 + 8 \b + 3 \b^2 \rb \rsb
\df \phi(\zwo) + \lsb \fr{8}{5 \,\zo} \lb 5 + 4 \b \rb 
- \fr{2\, \b\, \zwo^2}{5\, q^2\, \zo} \lb 5 + 2 \b \rb 
\rd \nonumber \\ \nonumber \\&-& \ld
\fr{\a\, \b\, \zwo^2}{\zo^3} \lb 1 + \b \rb \rsb\, \dt \phi(\zwo) 
+ \lsb \fr{8}{\zwo\, \zo} + \fr{4\, \b\, \zwo}{5\, q^2\, \zo} \lb 4 \a
\b -  5 \rb 
\rd \nonumber \\ \nonumber \\ &-& \ld
\fr{\b^2\,  \zwo}{5\, q^4\, \zo} \lb 5 \zwo^2 + 11 \zo^2 \rb  
-\fr{\a \,\b \,\zwo}{\zo^3} \lb 2 + \a \b \rb + \fr{\a\, \b^2\,
\zwo^3}{q^2\, 
\zo^3}  \rsb \dw \phi(\zwo) \label{eq:ap2}
\eea
\bea 
\xitw(\zo,\zwo)&=& \lsb 4 \b \lb {1 \over 3} + {\b \over 7} \rb 
+ \fr{2\, \b\, \zwo^2}{3\, \zo^2} \lb 1 + \b \rb \rsb \df \phi(\zwo)
+ \lsb \fr{8\, \b}{3\, \zwo} - \fr{10\, \b^2\, \zwo}{21 q^2} 
\rd \nonumber \\ \nonumber \\ &+& \ld
\fr{2\, \b \, \zwo}{3\, \zo^2} \lb 2 - 2\, \a - \a \b \rb 
-\fr{2\, \b^2 \, \zwo^3}{3\, q^2 \zo^2} \rsb \dt \phi(\zwo) 
+ \lsb \fr{4\,  \a\, \b^2}{3\, q^2} - \fr{2 \b^2 \zwo^2}{q^4} 
\rd \nonumber \\ \nonumber \\ &-& \ld
\fr{2\, \a\, \b}{3\, \zo^2} \lb 4 + \a \b \rb
+ \fr{2 \a \, \b^2\, \zwo^2}{q^2\, \zo^2} 
-\fr{2 \b^2 \zo^2}{3 q^4}
 \rsb \dw \phi(\zwo) \label{eq:ap3}
\eea
Equations (\ref{eq:f6}), (\ref{eq:f7}) and   (\ref{eq:f8}) 
can be written as 
\bea
\dt \phi(\zwo) &=& \fr{35}{\b^2 8} F_3(\zwo) + f_3(\zwo) A_3
\label{eq:ap4} \\
\dw \phi(\zwo) &=& \fr{35}{\b^2 8} F_2(\zwo) + f_2(\zwo) A_3 + A_2
\label{eq:ap5} \\
\df \phi(\zwo) &=& \fr{35}{\b^2 8} \xit4+ \fr{2 \zwo}{q^2} \dt
\phi(\zwo) \label{eq:ap6}
\eea

where the functions $f_2(\zwo)$ and $f_3(\zwo)$ are defined as
\bea
f_3(\zwo)&=&e^{(\zwo^2-z_L^2)/q^2}  \\
f_2(\zwo)&=&\int^{\zwo}_{z_L} dr  e^{(r^2-z_L^2)/q^2} 
\eea
and $F_2(\zwo)$ and $F_3(\zwo)$ are integrals of the observed
$\xitf(\zwo)$   
\bea 
F_3(\zwo)&=&  e^{\zwo^2/q^2}   \int^{\zwo}_{z_L} \, dy \, e^{-y^2/q^2}
\xitf(y)  \\
F_2(\zwo)&=&  \int^{\zwo}_{z_L} dr \,  e^{r^2/q^2} 
 \int^{r}_{z_L} dy \, e^{-y^2/q^2} \xitf(y) \,.
\eea

Using equations (\ref{eq:ap4}), (\ref{eq:ap5}) and (\ref{eq:ap6}) for
the derivatives of $\phi(\zwo)$ in equations (\ref{eq:ap1}),
(\ref{eq:ap2}) and (\ref{eq:ap3}), we obtain the following expressions
for $\xitz$, $\xito$ and $\xitw$ in terms of the observed $\xitf$ and
three parameters $\b$, $A_1$ and $A_2$.
\bea
\xitz(\zo,\zwo)&=& \lb {{-2\,\b}\over {{q^2}}} + 
     {{2\,\a \,{\b^2}}\over {3\,{q^2}}} + {4\over {{\zwo^2}}} - 
     {{{\b^2}\,{\zwo^2}}\over {{q^4}}} + {4\over {{{{ \zo}}^2}}} + 
     {{2\,\a \,\b}\over {3\,{{{ \zo}}^2}}} - 
     {{{{\a }^2}\,{\b^2}}\over {3\,{{{ \zo}}^2}}} - 
     {{2\,\b\,{\zwo^2}}\over {{q^2}\,{{{ \zo}}^2}}} + 
     {{\a \,{\b^2}\,{\zwo^2}}\over {{q^2}\,{{{ \zo}}^2}}}
\rd \nonumber \\ \nonumber \\&-& \ld 
     {{{\b^2}\,{{{ \zo}}^2}}\over {3\,{q^4}}} \rb
 \lsb  \,{ A_2}  \, + {f_2}(\zwo)  A_3 \rsb
    + \lb {4\over \zwo} + {{4\,\b}\over {3\,\zwo}} + {{2\,\zwo}\over
{{q^2}}} +  
        {{\b\,\zwo}\over {3\,{q^2}}} + {{{\b^2}\,\zwo}\over {3\,{q^2}}} + 
        {{4\,\zwo}\over {{{{ \zo}}^2}}} 
 \rd  \nonumber \\ \nonumber \\ &+& \ld   
        {{8\,\b\,\zwo}\over {3\,{{{ \zo}}^2}}} + 
        {{\a \,\b\,\zwo}\over {3\,{{{ \zo}}^2}}} - 
        {{\a \,{\b^2}\,\zwo}\over {3\,{{{ \zo}}^2}}} + 
        {{2\,{\zwo^3}}\over {{q^2}\,{{{ \zo}}^2}}} + 
        {{5\,\b\,{\zwo^3}}\over {3\,{q^2}\,{{{ \zo}}^2}}} + 
        {{{\b^2}\,{\zwo^3}}\over {3\,{q^2}\,{{{ \zo}}^2}}} \rb \,{
f_3}(\zwo) 
       A_3
  \nonumber \\ \nonumber \\ &+& 
 35  \lb {\a \over {12\,{q^2}}} - {1 \over {4\,\b\,{q^2}}} + 
     {1\over {2\,{\b^2}\,{\zwo^2}}} - {{{\zwo^2}}\over {8\,{q^4}}} - 
     {{{{\a }^2}}\over {24\,{{{ \zo}}^2}}} + 
     {1\over {2\,{\b^2}\,{{{ \zo}}^2}}} + 
     {{\a }\over {12\,\b\,{{{ \zo}}^2}}} + 
     {{\a \,{\zwo^2}}\over {8\,{q^2}\,{{{ \zo}}^2}}} 
\rd   \nonumber \\ \nonumber \\ &-& \ld
     {{{\zwo^2}}\over {4\,\b\,{q^2}\,{{{ \zo}}^2}}} - 
     {{{{{ \zo}}^2}}\over {24\,{q^4}}} \rb \,{ F_2}(\zwo) 
+35  \lb {{1}\over {2\,{\b^2}\,\zwo}} + {{1}\over {6\,\b\,\zwo}} + 
     {{\zwo}\over {24\,{q^2}}} +
 {{\zwo}\over {4\,{\b^2}\,{q^2}}} 
\rd   \nonumber \\ \nonumber \\ &+& \ld
     {{\zwo}\over {24\,\b\,{q^2}}} - 
     {{\a \,\zwo}\over {24\,{{{ \zo}}^2}}} + 
     {{\zwo}\over {2\,{\b^2}\,{{{ \zo}}^2}}} + 
     {{\zwo}\over {3\,\b\,{{{ \zo}}^2}}} + 
     {{\a \,\zwo}\over {24\,\b\,{{{ \zo}}^2}}} + 
     {{{\zwo^3}}\over {24\,{q^2}\,{{{ \zo}}^2}}} + 
     {{{\zwo^3}}\over {4\,{\b^2}\,{q^2}\,{{{ \zo}}^2}}} 
\rd   \nonumber \\ \nonumber \\ &+& \ld
     {{5\,{\zwo^3}}\over {24\,\b\,{q^2}\,{{{ \zo}}^2}}} \rb \,{F_3}(\zwo) 
	\nonumber \\ \nonumber \\ &+&
    \lb {7\over 8} + {{35}\over {8\,{\b^2}}} + {{35}\over {12\,\b}} + 
     {{35\,{\zwo^2}}\over {24\,{{{ \zo}}^2}}} + 
     {{35\,{\zwo^2}}\over {8\,{\b^2}\,{{{ \zo}}^2}}} + 
     {{35\,{\zwo^2}}\over {6\,\b\,{{{ \zo}}^2}}} \rb \,{ \xitf}(\zwo)
\nonumber \\
\eea
\bea
\xito(\zo,\zwo)&=&\lb {{-2\,\a \,\b\,\zwo}\over {{{{ \zo}}^3}}} - 
     {{{{\a }^2}\,{\b^2}\,\zwo}\over {{{{ \zo}}^3}}} + 
     {{\a \,{\b^2}\,{\zwo^3}}\over {{q^2}\,{{{ \zo}}^3}}} + 
     {8\over {\zwo\,{ \zo}}} - {{4\,\b\,\zwo}\over {{q^2}\,{ \zo}}} + 
     {{16\,\a \,{\b^2}\,\zwo}\over {5\,{q^2}\,{ \zo}}} - 
\rd   \nonumber \\ \nonumber \\ &-& \ld
     {{{\b^2}\,{\zwo^3}}\over {{q^4}\,{ \zo}}}-
     {{11\,{\b^2}\,\zwo\,{ \zo}}\over {5\,{q^4}}} 
\rb \,  \lsb A_2 \,+ \,{f_2}(\zwo) A_3 \rsb  +  
     \lb -{{\a \,\b\,{\zwo^2}}\over {{{{ \zo}}^3}}} - 
        {{\a \,{\b^2}\,{\zwo^2}}\over {{{{ \zo}}^3}}} + 
        {8\over {{ \zo}}}
\rd  \nonumber \\ \nonumber \\ &+& \ld 
  {{32\,\b}\over {5\,{ \zo}}} + 
        {{4\,{\zwo^2}}\over {{q^2}\,{ \zo}}} + 
        {{22\,\b\,{\zwo^2}}\over {5\,{q^2}\,{ \zo}}} + 
        {{8\,{\b^2}\,{\zwo^2}}\over {5\,{q^2}\,{ \zo}}} \rb \,{ f_3}(\zwo)
      A_3 \, +
  \lb {{-35\,{{\a }^2}\,\zwo}\over {8\,{{{ \zo}}^3}}} - 
     {{35\,\a \,\zwo}\over {4\,\b\,{{{ \zo}}^3}}}
\rd  \nonumber \\ \nonumber \\ &+& \ld
     {{35\,\a \,{\zwo^3}}\over {8\,{q^2}\,{{{ \zo}}^3}}} + 
     {{35}\over {{\b^2}\,\zwo\,{ \zo}}} + 
     {{14\,\a \,\zwo}\over {{q^2}\,{ \zo}}} - 
     {{35\,\zwo}\over {2\,\b\,{q^2}\,{ \zo}}} - 
     {{35\,{\zwo^3}}\over {8\,{q^4}\,{ \zo}}} - 
     {{77\,\zwo\,{ \zo}}\over {8\,{q^4}}} \rb \,{ F_2}(\zwo)  
\nonumber \\ \nonumber \\ &+&	
     \lb {{-35\,\a \,{\zwo^2}}\over {8\,{{{ \zo}}^3}}} - 
     {{35\,\a \,{\zwo^2}}\over {8\,\b\,{{{ \zo}}^3}}} + 
     {{35}\over {{\b^2}\,{ \zo}}} + {{28}\over {\b\,{ \zo}}} + 
     {{7\,{\zwo^2}}\over {{q^2}\,{ \zo}}} + 
     {{35\,{\zwo^2}}\over {2\,{\b^2}\,{q^2}\,{ \zo}}} + 
     {{77\,{\zwo^2}}\over {4\,\b\,{q^2}\,{ \zo}}} \rb \,{ F_3}(\zwo)
\nonumber \\ \nonumber \\ &+&	
  \lb {{21\,\zwo}\over {4\,{ \zo}}} + {{35\,\zwo}\over {4\,{\b^2}\,{ \zo}}} + 
     {{14\,\zwo}\over {\b\,{ \zo}}} \rb \,{ \xitf}(\zwo)
\nonumber \\
\eea
\bea
\xitw(\zo,\zwo)&=& \lb {{4\,\a \,{\b^2}}\over {3\,{q^2}}} - 
     {{2\,{\b^2}\,{\zwo^2}}\over {{q^4}}} - 
     {{8\,\a \,\b}\over {3\,{{{ \zo}}^2}}} - 
     {{2\,{{\a }^2}\,{\b^2}}\over {3\,{{{ \zo}}^2}}} + 
     {{2\,\a \,{\b^2}\,{\zwo^2}}\over {{q^2}\,{{{ \zo}}^2}}} - 
     {{2\,{\b^2}\,{{{ \zo}}^2}}\over {3\,{q^4}}} \rb  \, 
\lsb  { A_2}
\rd \nonumber \\ \nonumber \\ &+& \ld 
  \,{ f_2}(\zwo) A_3 \rsb +
     \lb {{8\,\b}\over {3\,\zwo}} + {{8\,\b\,\zwo}\over {3\,{q^2}}} + 
        {{2\,{\b^2}\,\zwo}\over {3\,{q^2}}} + 
        {{4\,\b\,\zwo}\over {3\,{{{ \zo}}^2}}} - 
        {{4\,\a \,\b\,\zwo}\over {3\,{{{ \zo}}^2}}} - 
        {{2\,\a \,{\b^2}\,\zwo}\over {3\,{{{ \zo}}^2}}} 
\rd \nonumber \\ \nonumber \\ &+& \ld 
        {{4\,\b\,{\zwo^3}}\over {3\,{q^2}\,{{{ \zo}}^2}}}  +
        {{2\,{\b^2}\,{\zwo^3}}\over {3\,{q^2}\,{{{ \zo}}^2}}} \rb \,
      { f_3}(\zwo)  \,  { A_3} + 35
	\lb {{\a }\over {6\,{q^2}}} -
     {{{\zwo^2}}\over {4\,{q^4}}} - 
     {{{{\a }^2}}\over {12\,{{{ \zo}}^2}}} - 
     {{\a }\over {3\,\b\,{{{ \zo}}^2}}} + 
\rd \nonumber \\ \nonumber \\ &+& \ld
     {{\a \,{\zwo^2}}\over {4\,{q^2}\,{{{ \zo}}^2}}} -
     {{{{{ \zo}}^2}}\over {12\,{q^4}}} \rb \,{ F_2}(\zwo) \, + \, 35
   \lb {{1}\over {3\,\b\,\zwo}} + 
     {{\zwo}\over {12\,{q^2}}} + {{\zwo}\over {3\,\b\,{q^2}}} - 
     {{\a \,\zwo}\over {12\,{{{ \zo}}^2}}}
\rd \nonumber \\ \nonumber \\ &+& \ld 
     {{\zwo}\over {6\,\b\,{{{ \zo}}^2}}} -
     {{\a \,\zwo}\over {6\,\b\,{{{ \zo}}^2}}} + 
    {{{\zwo^3}}\over {12\,{q^2}\,{{{ \zo}}^2}}} + 
     {{{\zwo^3}}\over {6\,\b\,{q^2}\,{{{ \zo}}^2}}} \rb {F_3}(\zwo)
 \nonumber \\ \nonumber \\ &+&
{5\over 2}  \lb 1  + {{7}\over {3\,\b}} + 
     {{7\,{\zwo^2}}\over {6\,{{{ \zo}}^2}}} + 
     {{7\,{\zwo^2}}\over {6\,\b\,{{{ \zo}}^2}}} \rb \,{ \xitf}(\zwo)
\nonumber \\
\eea

\begin{figure}
\figurenum{2c}
\plotfiddle{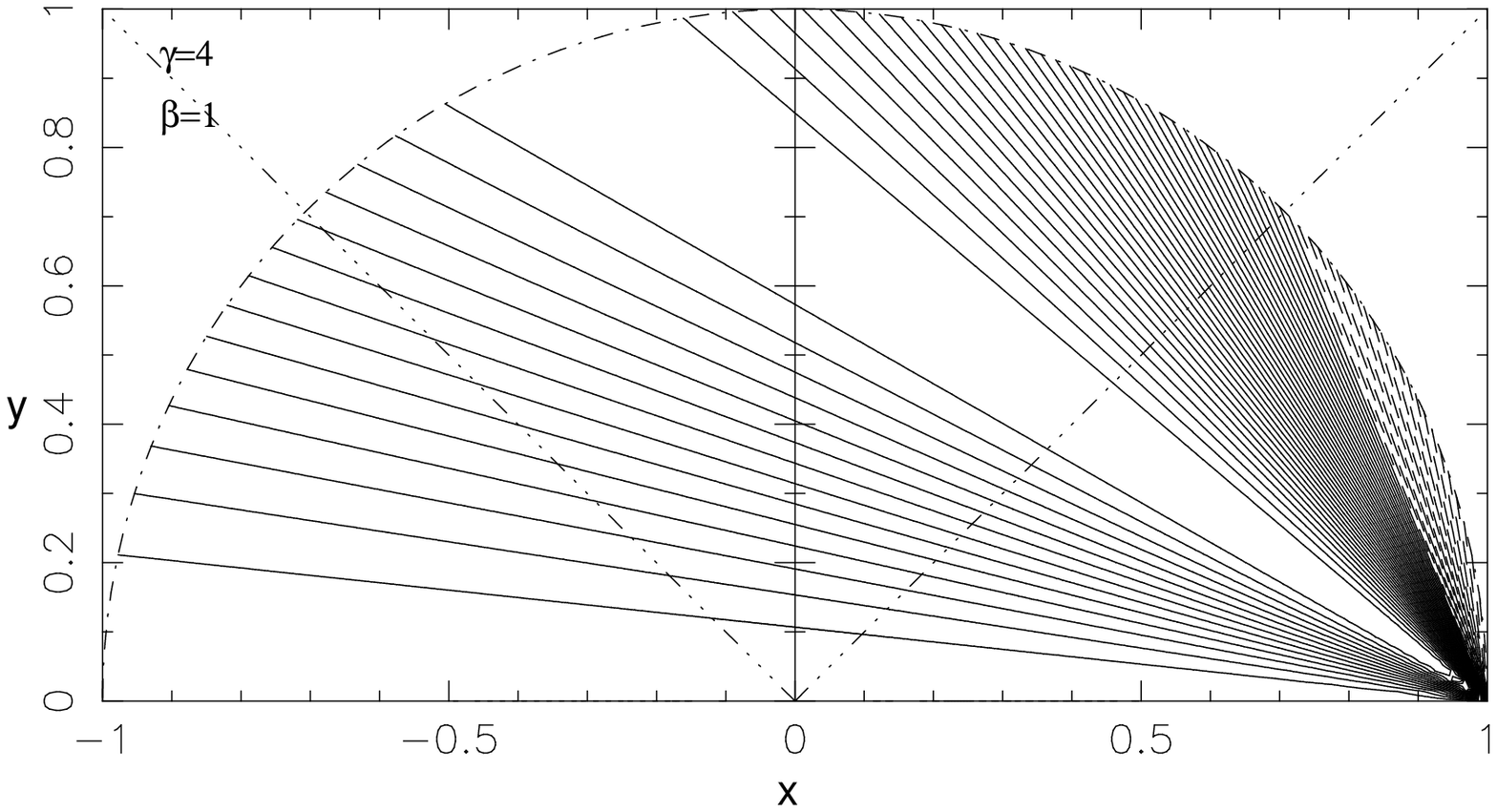}{3in}{0}{80}{80}{-240}{-200}
\caption{This shows contours of equal $w$ calculated using  the plane
parallel approximation plotted in the parameter
space corresponding to triangles of all possible shapes. 
The solid lines show positive values of $w$ and the dashed lines show
negative values of $w$.} 
\label{fig:cn4}
\end{figure}

\begin{figure}
\figurenum{2d}
\plotfiddle{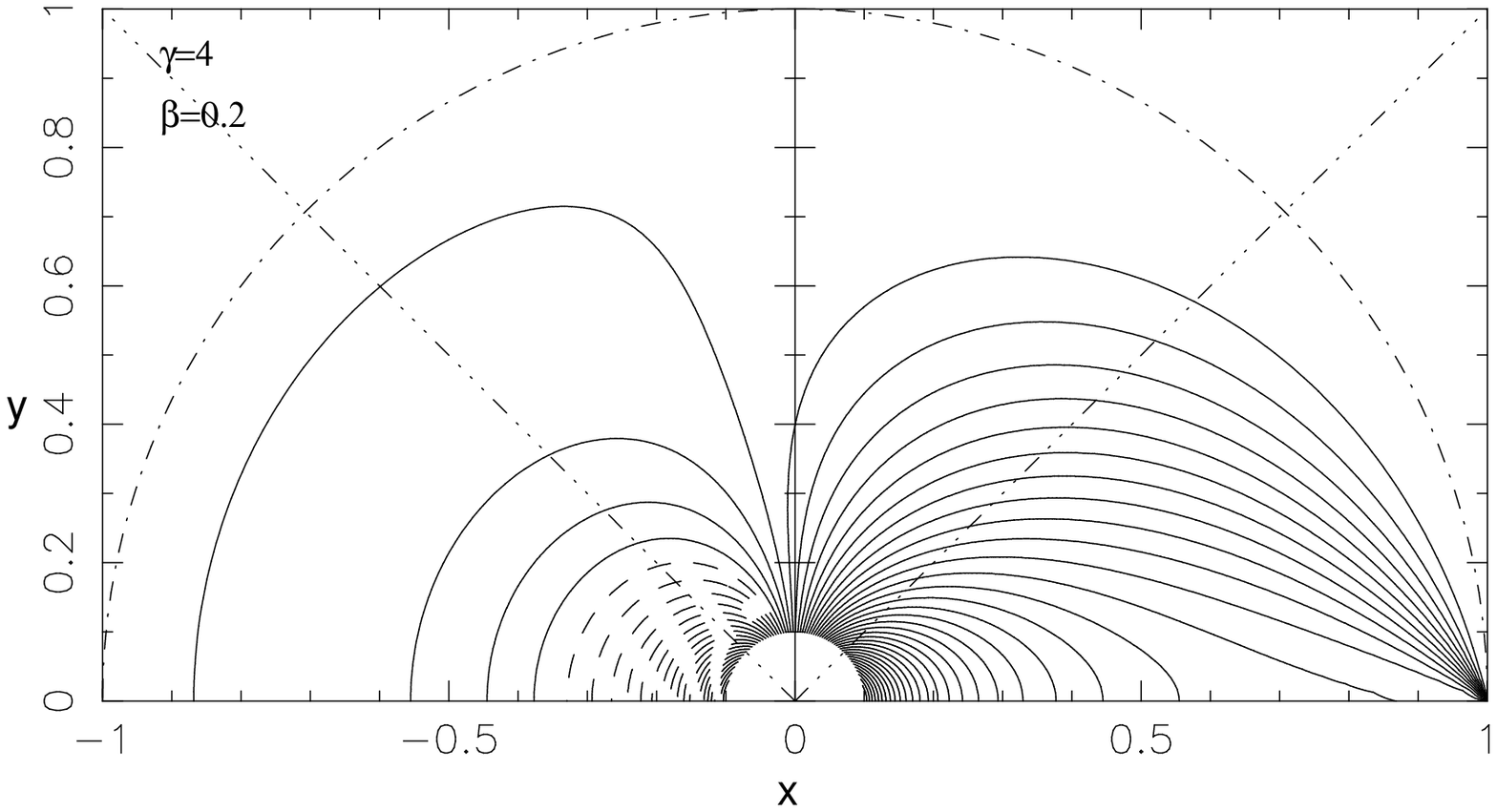}{3in}{0}{80}{80}{-240}{-200}
\caption{This shows contours of equal $w$ plotted in the parameter
space corresponding to triangles of all possible shapes. 
The solid lines show positive values of $w$ and the dashed lines show
negative values of $w$.} 
\label{fig:cn5}
\end{figure}

\begin{figure}
\figurenum{3a}
\plotfiddle{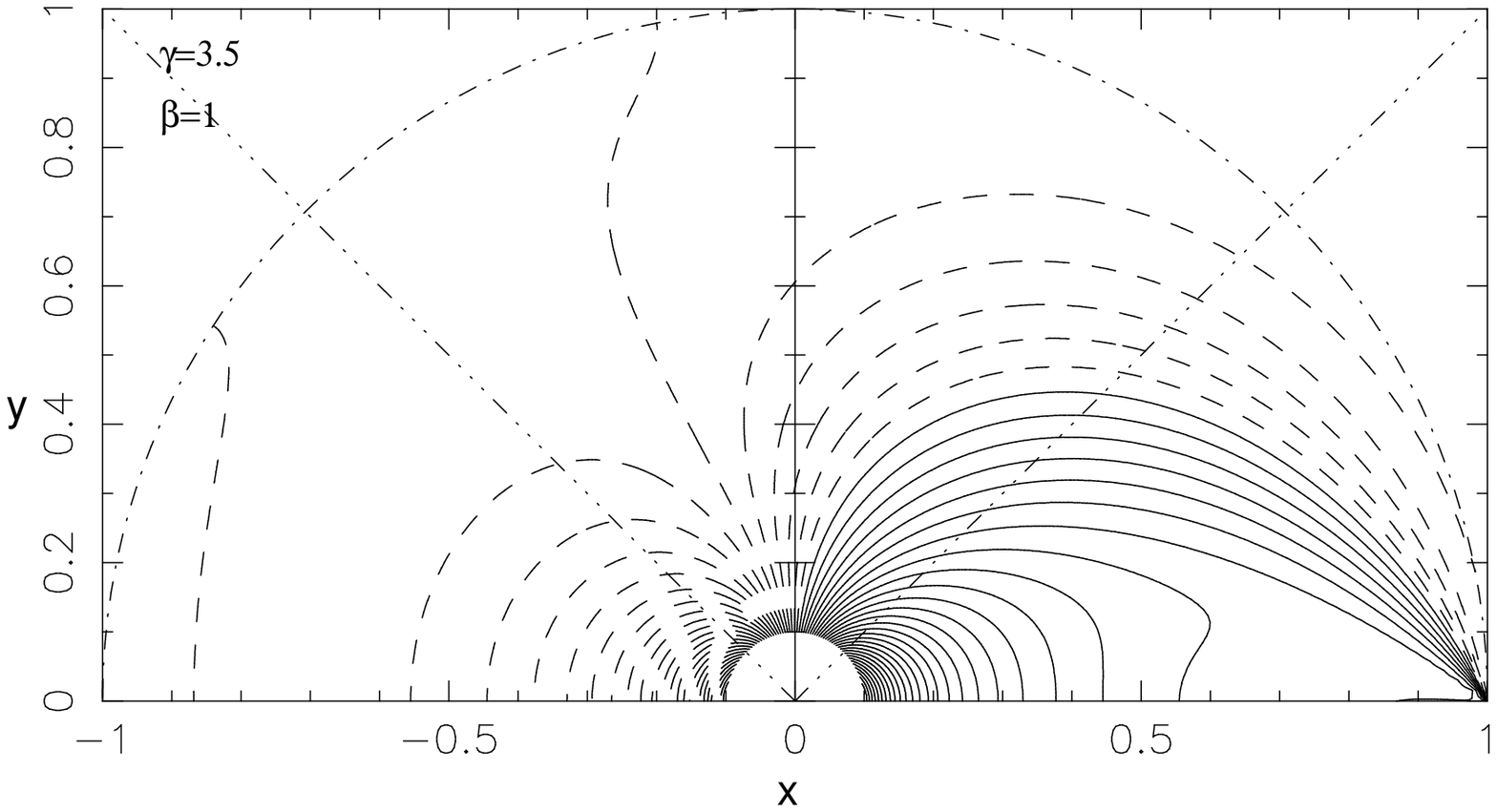}{3in}{0}{80}{80}{-240}{-200}
\caption{This shows contours of equal $w$ plotted in the parameter
space corresponding to triangles of all possible shapes. 
The solid lines show positive values of $w$ and the dashed lines show
negative values of $w$.} 
\label{fig:cn2}
\end{figure}

\begin{figure}
\figurenum{3b}
\plotfiddle{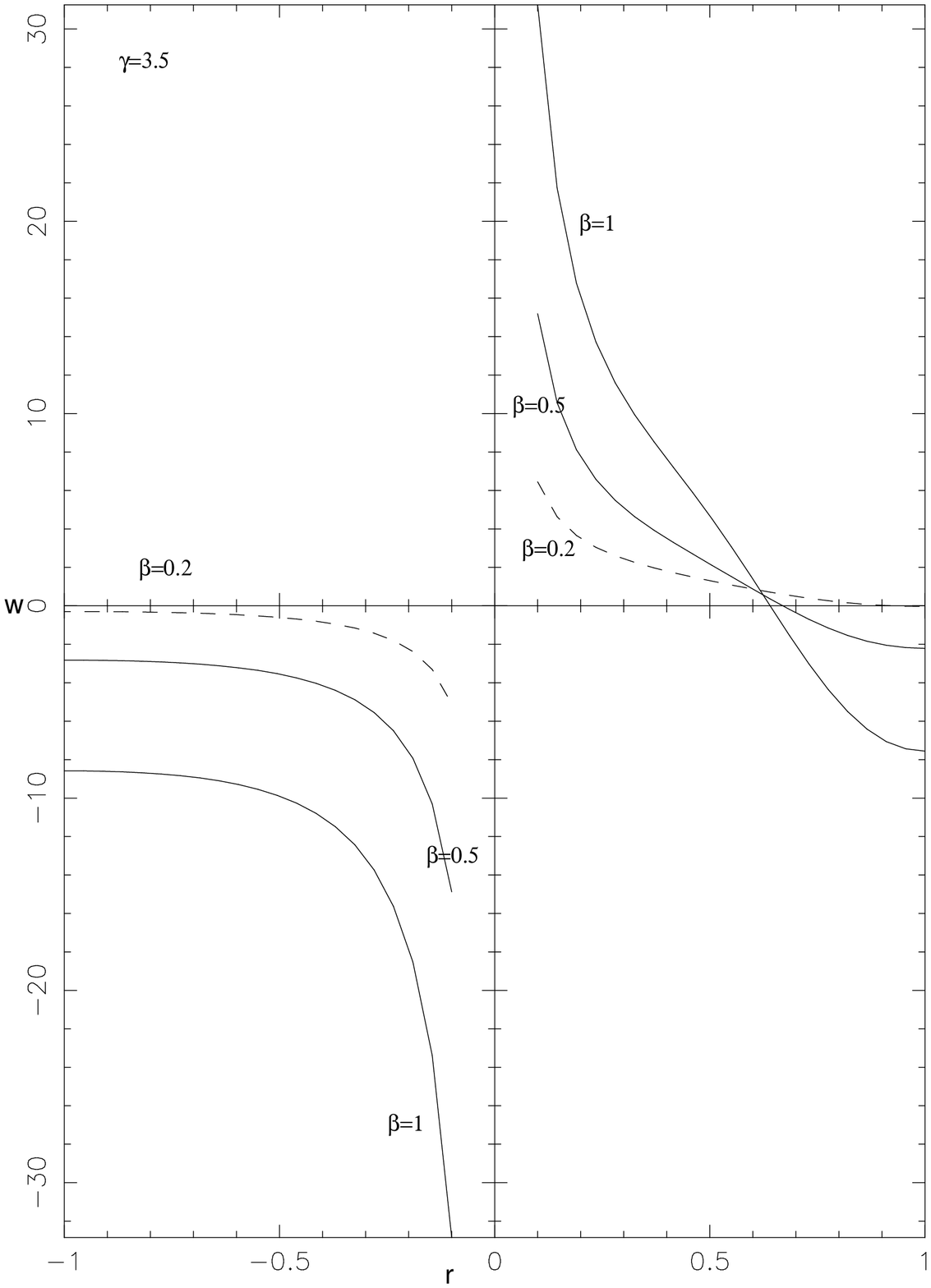}{3in}{0}{40}{40}{-80}{-30}
\caption{This shows $w$ along a cross-section at $45^{\circ}$ through figure
\ref{fig:cn2}.}
\label{fig:ct2}
\end{figure}

\begin{figure}
\figurenum{4a}
\plotfiddle{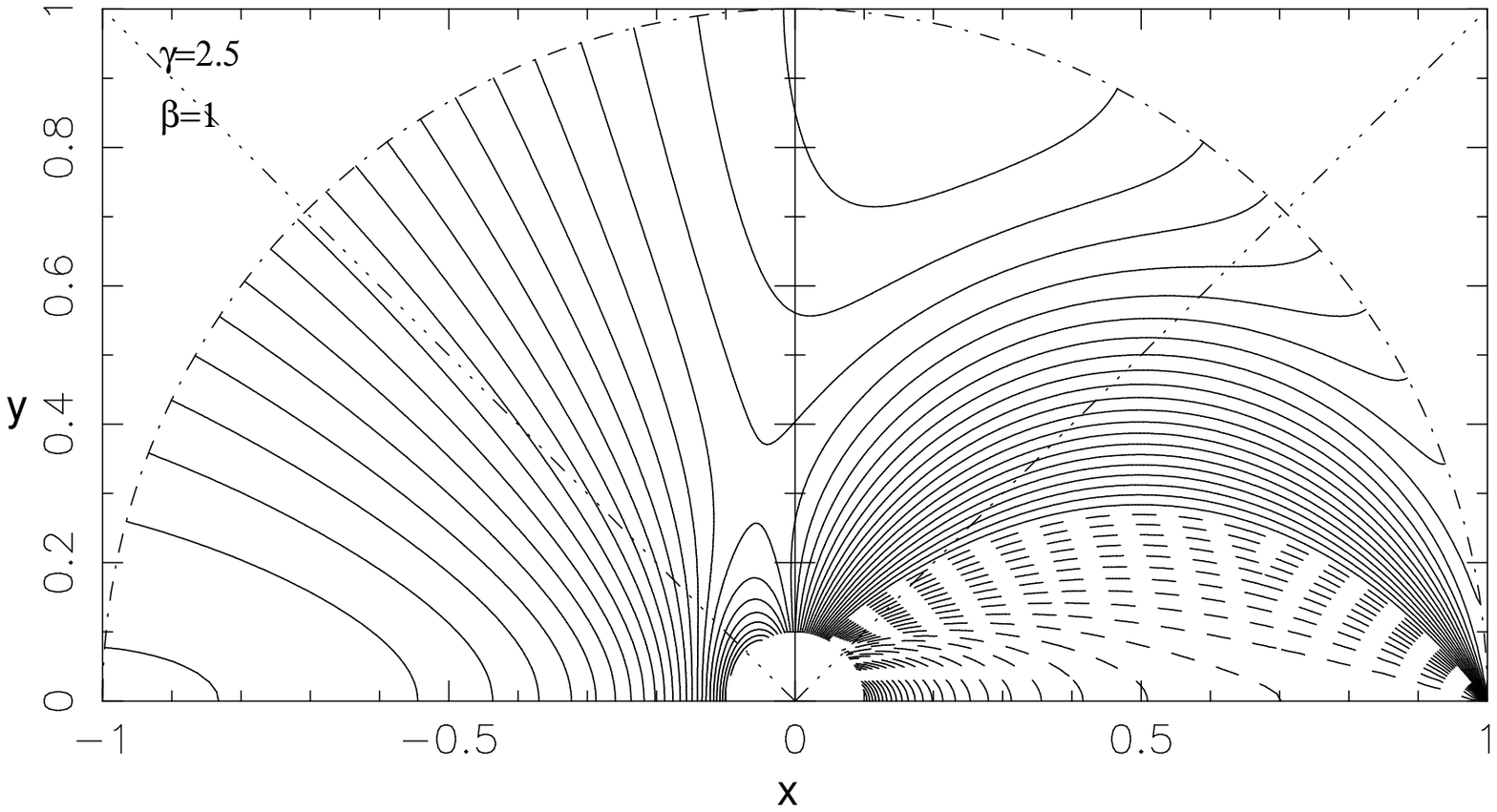}{3in}{0}{80}{80}{-240}{-200}
\caption{This shows contours of equal $w$ plotted in the parameter
space corresponding to triangles of all possible shapes. 
The solid lines show positive values of $w$ and the dashed lines show
negative values of $w$.} 
\label{fig:cn3}
\end{figure}

\begin{figure}
\figurenum{4b}
\plotfiddle{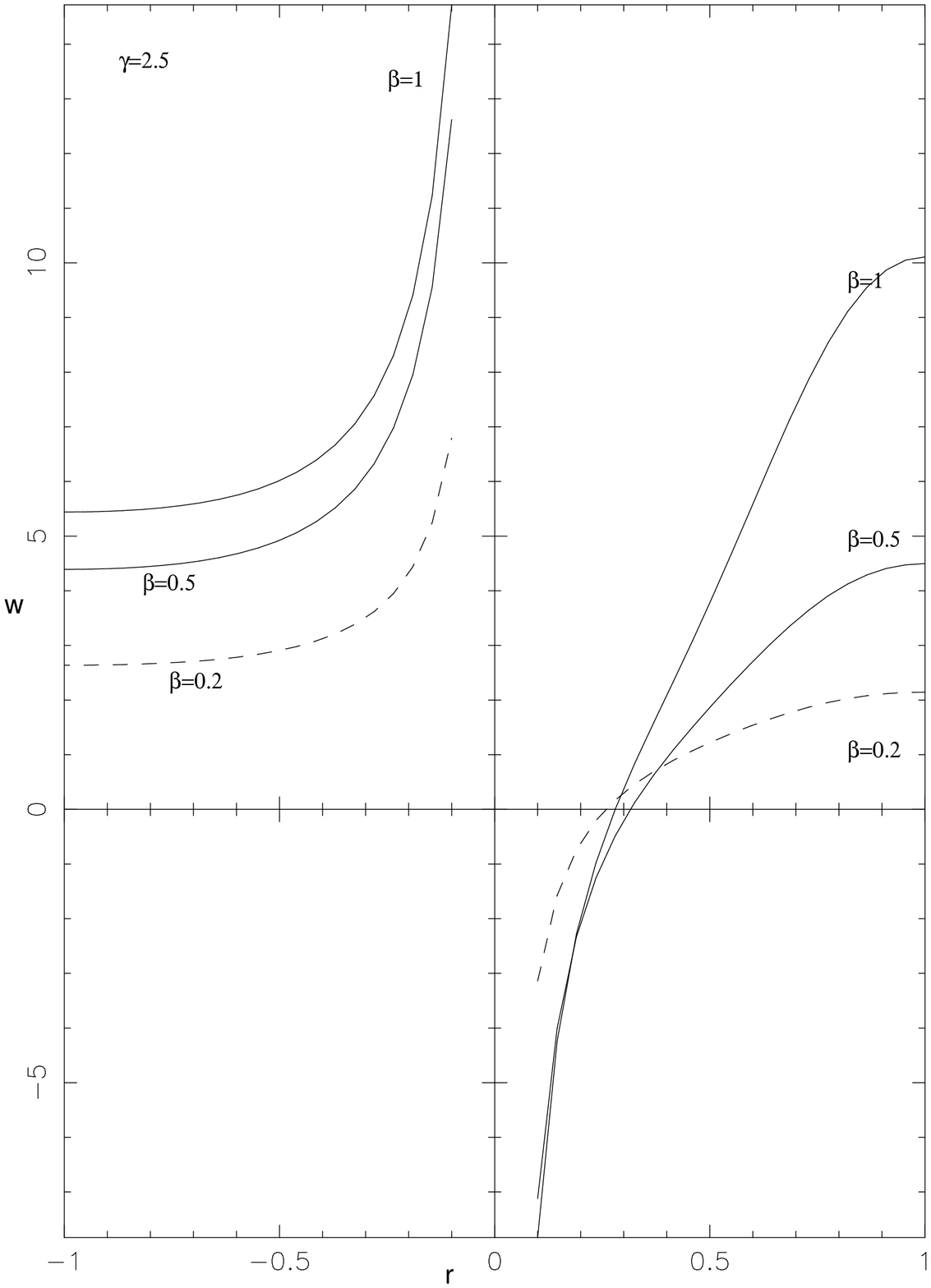}{3in}{0}{40}{40}{-80}{-30}
\caption{This shows $w$ along a cross-section at $45^{\circ}$ through figure
\ref{fig:cn3}.}
\label{fig:ct3}
\end{figure}

\begin{figure}
\figurenum{5}
\label{fig:f1}
\plotfiddle{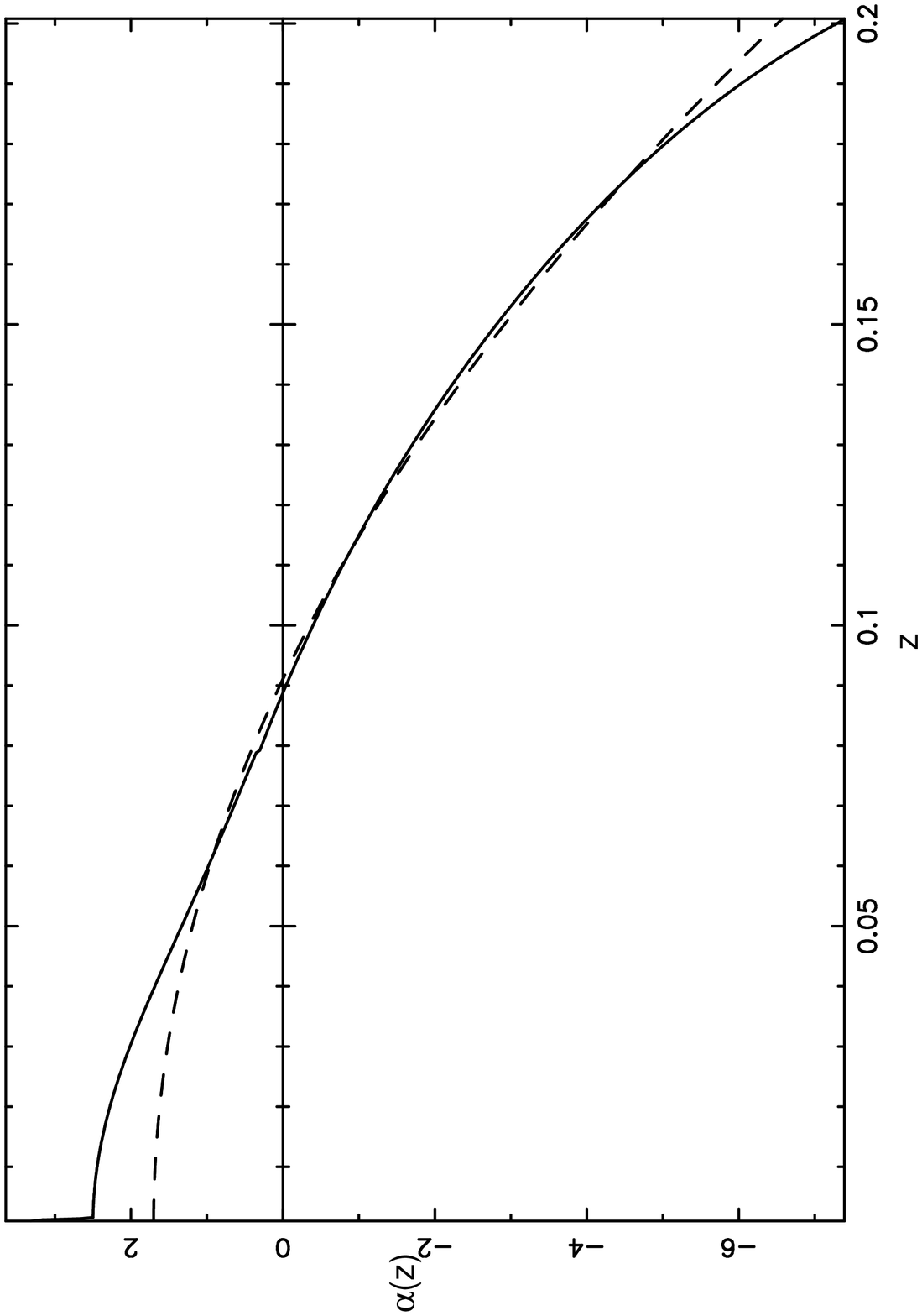}{3in}{-90}{50}{50}{-170}{270}
\caption{This solid curve shows the selection function $\Phi(z)$
for the N112 sample of LCRS. The dashed-dotted curve shows the fit
using equation (\ref{eq:f2}). The dashed line shows $n(<z)$ - the
fraction of galaxies that are expected to lie below a redshift $z$.}
\end{figure} 

\begin{figure}
\figurenum{6}
\plotfiddle{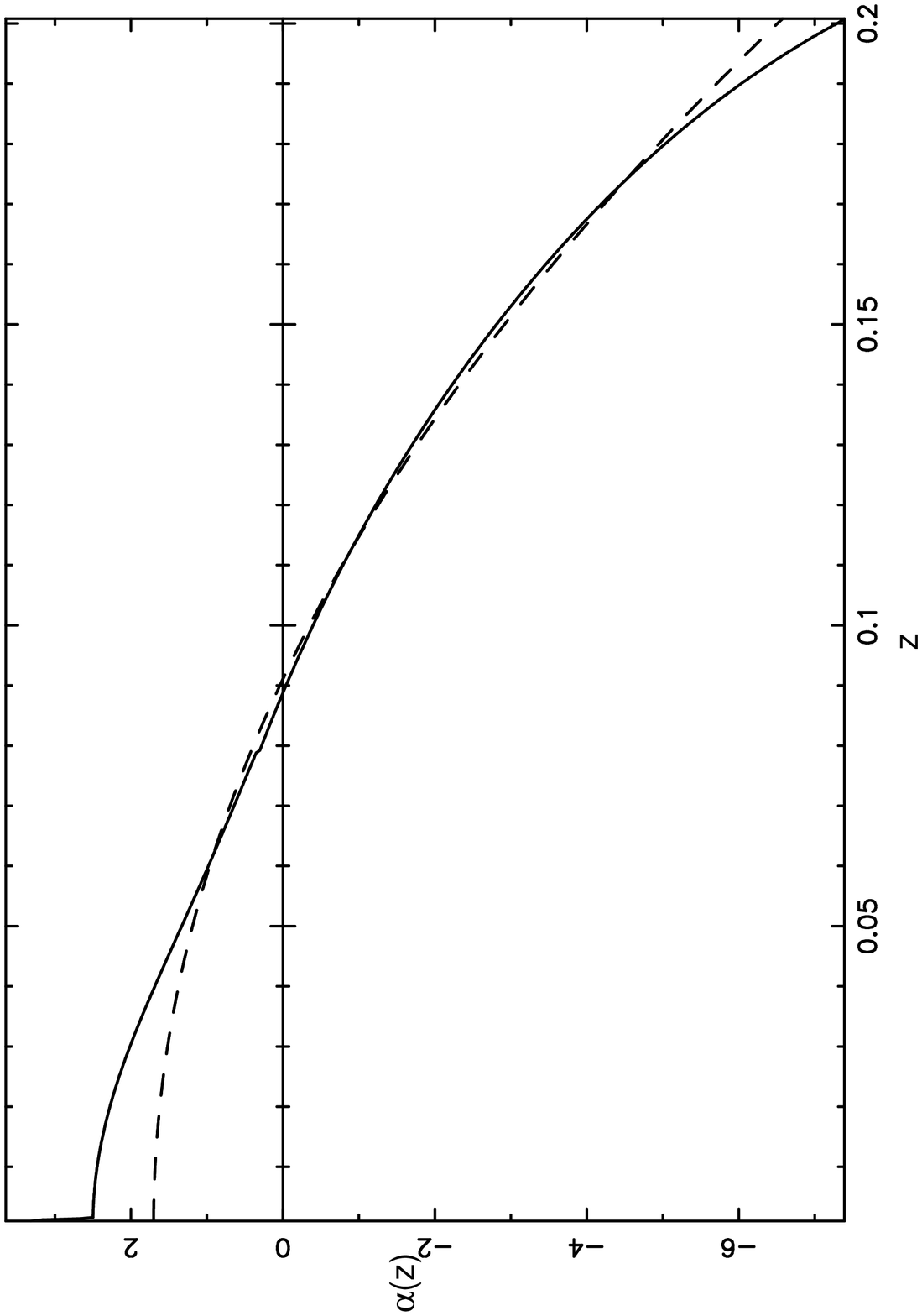}{3in}{-90}{50}{50}{-170}{270}
\caption{This solid curve shows $\a(z)$  for the N112 sample of
LCRS. The dashed-dotted curve shows the fit using equation
(\ref{eq:f3}).}
\label{fig:f2}
\end{figure} 

\ed